\documentclass[aps,reprint,amsmath,amssymb]{revtex4-2}

\usepackage{graphicx}
\usepackage{dcolumn}
\usepackage{bm}
\usepackage{subfigure}
\usepackage{float}
\usepackage{amssymb}
\usepackage{color}
\usepackage{mathrsfs}
\usepackage{amstext}

\begin{document}


\title{High Precision Multi-parameter Weak Measurement with Hermite-Gaussian Pointer}


\author{Binke Xia}
\author{Jingzheng Huang}
\email{jzhuang1983@sjtu.edu.cn}
\author{Chen Fang}
\author{Hongjing Li}
\author{Guihua Zeng}
\email{ghzeng@sjtu.edu.cn}
\affiliation{State Key Laboratory of Advanced Optical Communication Systems and Networks, Center of Quantum Sensing and Information Processing, Shanghai Jiao Tong University, Shanghai 200240, China}


\date{\today}

\begin{abstract}
The weak value amplification technique has been proved useful for precision metrology in both theory and experiment. To explore the ultimate performance of weak value amplification for multi-parameter estimation, we investigate a general weak measurement formalism with assistance of high-order Hermite-Gaussian pointer and quantum Fisher information matrix. Theoretical analysis shows that the ultimate precision of our scheme is improved by a factor of square root of 2n+1, where n is the order of Hermite-Gaussian mode. Moreover, the parameters' estimation precision can approach the precision limit with maximum likelihood estimation method and homodyne method. We have also given a proof-of-principle experimental setup to validate the H-G pointer theory and explore its potential applications in precision metrology.
\end{abstract}


\maketitle

\section{Introduction}
The concepts of weak measurement (WM) and weak value amplification (WVA), which were proposed by Aharonov, Albert, and Vaidman\cite{PhysRevLett.60.1351}, have nowadays been developed to be an important technique for precision metrology\cite{PhysRevLett.66.1107,Hosten787,PhysRevLett.102.173601,PhysRevLett.114.210801,PhysRevA.97.063818,Fang_2016,PhysRevA.97.063853}.
While most previous works mainly focus on single parameter estimation\cite{PhysRevLett.102.173601,PhysRevLett.107.133603,nphys4040,doi:10.1063/1.5027117,PhysRevLett.105.010405,PhysRevLett.111.033604,RevModPhys.86.307}, recent progresses were made to extend the WVA technique for multi-parameter estimation\cite{Dziewior2881,PhysRevLett.122.123603,2018arXiv181108046H}. For instance, in\cite{PhysRevLett.122.123603}, authors extended the WVA formalism to the simultaneous measurement of multiple parameters in an optical focused vector beam scatterometry experiment, the resolutions are improved over 1000 times.

In the WVA formalism, two parties labeled by "system" and "pointer" are weakly interacted through a coupling coefficient which relates to the unknown parameter to be measured. By performing suitable pre- and post-selection on the system, the information of the parameter can be extracted from the significant changes on the pointer state\cite{RevModPhys.86.307,PhysRevX.4.011031,2019arXiv190106831M}. In this work, we extend this formalism by considering a weak interaction process with arbitrary coupling parameters. In particular we concern on the scenario with two unknown parameters, spatial displacement and its conjugated momentum kick.

In a conventional scheme, the initial pointer state is chosen in Gaussian distribution\cite{PhysRevA.41.11,PhysRevA.81.033813}. To explore the ultimate performance of WVA, a natural question arises: can employing pointer states in high-order Hermite-Gaussian (H-G) distributions be benefit in achieving higher precision? To analyze above weak measurement scenario involving a two-level state system and a pointer in high-order Hermite-Gaussian distribution, we employ the quantum Fisher information matrix (QFIM) as the figure of merit\cite{Helstrom1969,holevo2011probabilistic}. Our results show that H-G pointer states take advantages over the fundamental Gaussian state in improving the precision limit according to the Cram$\mathrm{\acute{e}}$r-Rao theorem\cite{DEMKOWICZDOBRZANSKI2015345}.  For every single parameter, the quantum Fisher information can be increased by a factor of $2n+1$, where $n$ is the order of the H-G distribution.

But the practical precision is restricted by the measurement strategies on final pointer state. Thus we propose two practical methods, namely the maximum likelihood estimation method and the homodyne detection method, that can approach the precision limit.

The new findings have many immediate applications. As an example, we propose an experimental setup to simultaneously measure ultra-small object displacement and tilt by using $n-th$ order H-G mode laser and optical homodyne detection. In principle, the precision of our scheme is by a factor of $\sqrt{2n+1}$ higher than that of a conventional WVA scheme with fundamental Gaussian pointer.

This paper is organized as follows. A general theory of multi-parameter weak measurement is established in Sec.\ref{sec:2}, and high-order H-G pointer is employed for weak measurement scenario with spatial displacement and momentum kick based on our theoretical framework. Two practical methods approaching the precision limit are proposed in Sec.\ref{sec:3}, and a proof-of-principle experimental setup is presented in Sec.\ref{sec:4}. Finally, discussions and conclusions are summarized in Sec.\ref{sec:5}.

\section{Multi-parameter Weak Measurement Process} \label{sec:2}
\subsection{Measurement Process}
\begin{figure}[h]
	\centering
	\includegraphics[scale=0.4]{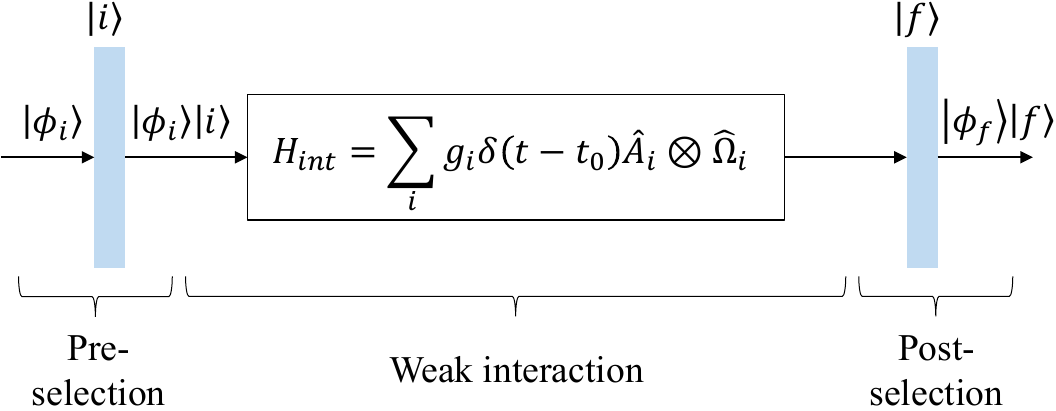}
	\caption{\label{fig:2-1} Multi-parameter weak measurement process.}
\end{figure}

First, we establish a theoretical framework for multiple parameter weak measurement. A standard weak measurement process consists of three parts: pre-selection, weak interaction and post-selection, as shown in Fig\ref{fig:2-1}.

We denote the initial pointer state as $|\phi_{i}\rangle$. Pre-selection couples pointer with pre-selection quantum state $|i\rangle$ in product state $|\phi_{i}\rangle|i\rangle$. The product state evolves to an entangled state after the weak interaction. We denote the unitary evolution operator as $\hat{U} = e^{-i\int H_{int}dt}$ (we adopt $\hbar=1$ here and hereafter), where $H_{int}$ is the interaction Hamiltonian. Post-selection projects the entangled state to $|\phi_{f}\rangle|f\rangle$, and the measurement information could be read out from the final pointer state $|\phi_{f}\rangle$. For multi-parameter weak measurement, the interaction Hamiltonian is denoted as
\begin{equation}
	H_{int} = \sum_{i}g_{i}\delta(t-t_{0})\hat{A}_{i}\otimes\hat{\Omega}_{i} \label{eq:sup2-1}
\end{equation}
where $g_{i}$ is the $i$-th coupling coefficient, i.e. the $i$-th unknown parameter to be measured. $g_{i}\ll 1$ is sufficient because of weak measurement restriction, and we denote all unknown parameters as a vector $\mathbf{g}=(g_{1},g_{2},...,g_{i},...)$. $\hat{A}_{i}$ is the corresponding measurement operator on quantum system, and $\hat{\Omega}_{i}$ is the corresponding general translation operator on pointer. Thus the unitary evolution operator $\hat{U}=e^{-i\sum_{i}g_{i}\hat{A}_{i}\hat{\Omega}_{i}}$. Then the final state of pointer is given by:
\begin{eqnarray}
	|\phi_{f}\rangle &&= \langle f|\hat{U}|i\rangle = \langle f|e^{-i\sum_{i}g_{i}\hat{A}_{i}\hat{\Omega}_{i}}|i\rangle|\phi_{i}\rangle \nonumber \\
	&&\approx \langle f|i\rangle(1-i\sum_{i}g_{i}A_{wi}\hat{\Omega}_{i})|\phi_{i}\rangle \label{eq:sup2-2}
\end{eqnarray}
where $A_{wi} = \langle f|\hat{A}_{i}|i\rangle/\langle f|i\rangle$ is the $i$-th weak value, the formulation here is remained in $O(\mathbf{g})$. Only take the first order of $O(\mathbf{g})$ into account, the normalized factor can be approximately given as $\langle f|i\rangle$. Thus the normalized final state is
\begin{equation}
	|\phi_{f}\rangle = (1-i\sum_{i}g_{i}A_{wi}\hat{\Omega}_{i})|\phi_{i}\rangle \label{eq:sup2-3}
\end{equation}

For simplicity, we investigate a common weak measurement scenario which focus on spatial displacement and conjugated momentum kick of pointer. We denote the spatial displacement as $d$ and the momentum kick as $k$. Thus the unknown parameter vector is $\mathbf{g}=(d,k)$. The corresponding translation operator of $d$ and $k$ are $\hat{p}$ and $\hat{x}$ separately. Thus the final state in this scenario is
\begin{equation}
	|\phi_{f}\rangle = (1-iA_{w}d\hat{p}-iA_{w}k\hat{x})|\phi_{i}\rangle \label{eq:sup2-4}
\end{equation}

In the conventional scheme, the initial pointer state is chosen in Gaussian distribution. To improve the measurement precision, we employ  a $n$-th order Hermite-Gaussian pointer in our scheme:
\begin{gather}
	|\phi_{i}\rangle = |\phi_{n}\rangle = \int_{-\infty}^{\infty}\,\mathrm{d}x\phi_{n}(x)|x\rangle \label{eq:2-1}\\
	\phi_{n}(x) = (\frac{1}{2\pi\sigma^{2}})^{\frac{1}{4}}\frac{1}{\sqrt{2^{n}n!}}H_{n}(\frac{x}{\sqrt{2}\sigma})e^{\frac{-x^{2}}{4\sigma^{2}}} \label{eq:2-2}
\end{gather}
where $H_{n}$ is the n-order Hermite polynomial. $\phi_{n}(x)=\langle x|\phi_{i}\rangle$ is the spatial distribution of pointer. In the $x$-representation, $\hat{p}$ is $-i\partial_{x}$, $\hat{x}$ is $x$. For the n-order Hermite-Gaussian pointer amplitude $\phi_{n}(x)$, we have the following relations: $\partial_{x}\phi_{n}(x)=(1/2\sigma)[\sqrt{n}\phi_{n-1}(x)-\sqrt{n+1}\phi_{n+1}(x)]$ and $x\phi_{n}(x)=\sigma[\sqrt{n}\phi_{n-1}(x)+\sqrt{n+1}\phi_{n+1}(x)]$. Plugging these into Eq.(\ref{eq:sup2-4}), the final state of pointer can be reformulated as:
\begin{eqnarray}
|\phi_{f}\rangle =&& |\phi_{n}\rangle-(\frac{A_{w}}{2\sigma}d+iA_{w}\sigma k)\sqrt{n}|\phi_{n-1}\rangle \nonumber \\
&&+(\frac{A_{w}}{2\sigma}d-iA_{w}\sigma k)\sqrt{n+1}|\phi_{n+1}\rangle \label{eq:2-5}
\end{eqnarray}

We denote the $n$-th order H-G pointer state as $|\phi_{n}\rangle$. According to the orthogonality of Hermite polynomials, we have $\langle \phi_{m}|\phi_{n}\rangle=\delta_{mn}$. Thus the final state $|\phi_{f}\rangle$ we derived is a linear superposition of orthogonal states $|\phi_{n}\rangle$ and its neighbor mode $|\phi_{n-1}\rangle$ and $|\phi_{n+1}\rangle$.(Here and hereafter we treat notation $|\phi_{f}\rangle$ as the normalized final pointer state.)

\subsection{Precision Limit of Multi-parameter Estimation}
In this part, we employ the quantum Fisher information matrix (QFIM) as the figure of merit to analyze a multi-parameter weak measurement scenario. The inverse of QFIM, which is the quantum Cram$\mathrm{\acute{e}}$r-Rao bound (QCRB), implies the ultimate estimation precision of final pointer state $|\phi_{f}\rangle$ about parameter vector $\mathbf{g}$.

We denote the quantum Fisher information as $\mathbb{Q}$, its component
\begin{equation}
	[\mathbb{Q}]_{ij}=\mathrm{Tr}\left[\rho_{\mathbf{g}}\frac{\hat{L}_{i}\hat{L}_{j}+\hat{L}_{j}\hat{L}_{i}}{2}\right] \label{eq:2-6}
\end{equation}
where $\rho_{\mathbf{g}}=|\phi_{f}\rangle\langle\phi_{f}|$ is the density matrix for final pointer state $|\phi_{f}\rangle$. We denote $|\partial_{g_{i}}\phi_{f}\rangle$ as shorthand $|\partial_{i}\phi_{f}\rangle$. $L_{i}$ is the Symmetric Logarithmic Derivative (SLD) with respect to parameter $g_{i}$. $L_{i}=2(|\partial_{i}\phi_{f}\rangle\langle\phi_{f}|+|\phi_{f}\rangle\langle\partial_{i}\phi_{f}|)$ for pure state $|\phi_{f}\rangle$. Thus $[\mathbb{Q}]_{ij}$ in Eq.(\ref{eq:2-6}) is given by:
\begin{eqnarray}
	[\mathbb{Q}]_{ij}=&&2(\langle\partial_{i}\phi_{f}|\partial_{j}\phi_{f}\rangle+\langle\partial_{j}\phi_{f}|\partial_{i}\phi_{f}\rangle) \nonumber\\
	&&-4\langle\partial_{i}\phi_{f}|\phi_{f}\rangle\langle\phi_{f}|\partial_{j}\phi_{f}\rangle \label{eq:2-7}
\end{eqnarray}
By using Eq.(\ref{eq:sup2-3}), the $i$-th row and $j$-th column element in QFIM is calculated as:
\begin{equation}
	[\mathbb{Q}]_{ij} = 4\mathrm{Re}(A_{wi}^{*}A_{wj}\langle\phi_{i}|\hat{\Omega}_{i}\hat{\Omega}_{j}|\phi_{i}\rangle) \label{eq:sup2-5}
\end{equation}
where only the constant term is remained because of $g_{i}\ll 1 \,(i=1,2,\cdots)$.

In our weak measurement scheme, we focus on two common parameters, the spatial displacement $d$ and its conjugated momentum kick $k$. Thus the translation operator $\hat{\Omega}_{1}=\hat{p}$, $\hat{\Omega}_{2}=\hat{x}$. And we assume that measurement $\hat{A}_{1}=\hat{A}_{2}=\hat{A}$. Here we employ $n$-th order H-G pointer, which is $|\phi_{i}\rangle=|\phi_{n}\rangle$. By using Eq.(\ref{eq:sup2-5}), the quantum Fisher information matrix of parameter $\mathbf{g}=(d,k)$ is calculated as:
\begin{equation}
	\label{eq:2-8}
	\mathbb{Q} = \left(
	\begin{array}{cc}
		(2n+1)|A_{w}|^{2}\sigma^{-2} & 0\\
		0 & 4(2n+1)|A_{w}|^{2}\sigma^{2}
	\end{array}
\right)
\end{equation}
Considering $N$ times independent same trials, the $N$ time quantum Fisher information matrix is $\mathbb{Q}_{N}=N\mathbb{Q}$. In the standard weak measurement frame, the effective samples reduce to $N'=P_{s}N$ because of the successful probability $P_{s}=|\langle f|i\rangle|^{2}$ in post-selection. Thus the $N'$ time quantum Fisher information matrix is
\begin{equation}
	\label{eq:2-9}
	\mathbb{Q}_{N'} = |\langle f|i\rangle|^{2}N(2n+1)|A_{w}|^{2}\left(
	\begin{array}{cc}
		\sigma^{-2} & 0\\
		0 & 4\sigma^{2}
	\end{array}
\right)
\end{equation} 
The QCRB of parameter vector $\mathbf{g}$ is $\mathbb{Q}_{N'}^{-1}$. Thus for parameters $d$ and $k$, the QCRB is given as:
\begin{equation}
	\label{eq:sup2-6}
	\mathbb{Q}_{N'}^{-1} = \frac{1}{4|\langle f|i\rangle|^{2}N(2n+1)|A_{w}|^{2}}\left(
	\begin{array}{cc}
		4\sigma^{2} & 0\\
		0 & \sigma^{-2}
	\end{array}
\right)
\end{equation} 
which gives the low bound of estimation covariance matrix about parameter $\mathbf{g}=(d,k)$, i.e. $\mathbb{C}_{N'}(\mathbf{g})\ge\mathbb{Q}_{N'}^{-1}(\mathbf{g})$. Thus from Eq.(\ref{eq:sup2-6}), we conclude that the measurement precision limit of parameter $\mathbf{g}=(d,k)$ can be improved by factor $2n+1$, where $n$ is the mode order of pointer. The merit is significant by applying high order H-G pointer to weak measurement.

\paragraph*{Numerical results} We using Eq.(\ref{eq:2-5}) to derive an approximation analytic result of QFIM above, and get some interesting conclusions. Here, we give a numerical simulation for QFI about $d$ and $k$ without approximation, which are the top left and lower right elements of QFIM. We expanded Eq.(\ref{eq:sup2-4}) and remained one order for approximation results before. Now, we do not expand the expression of final state, and calculated the results of QFIM directly. Though the analytic result without approximation is almost unable to give, the numerical result is a competent evidence to verify our approximation result above.
In our numerical simulation, we choose pre-selection $|i\rangle=\frac{1}{\sqrt{2}}(|H\rangle+|V\rangle)$ and post-selection $|f\rangle=e^{i\frac{\varepsilon}{2}}\cos (\frac{\pi}{4}+\frac{\varepsilon}{2})|H\rangle-e^{-i\frac{\varepsilon}{2}}\sin (\frac{\pi}{4}+\frac{\varepsilon}{2})|V\rangle$, where $\varepsilon=0.01$. Operator $\hat{A}=|H\rangle\langle H|-|V\rangle\langle V|$ is Pauli operator in this simulation. And the parameter $\sigma$ in pointer's transverse spatial distribution is set to 1.
\begin{figure}[h]
	\centering
	\subfigure[Spatial displacement]{
		\begin{minipage}{0.8\linewidth}
			\centering  
			\includegraphics[scale=0.37]{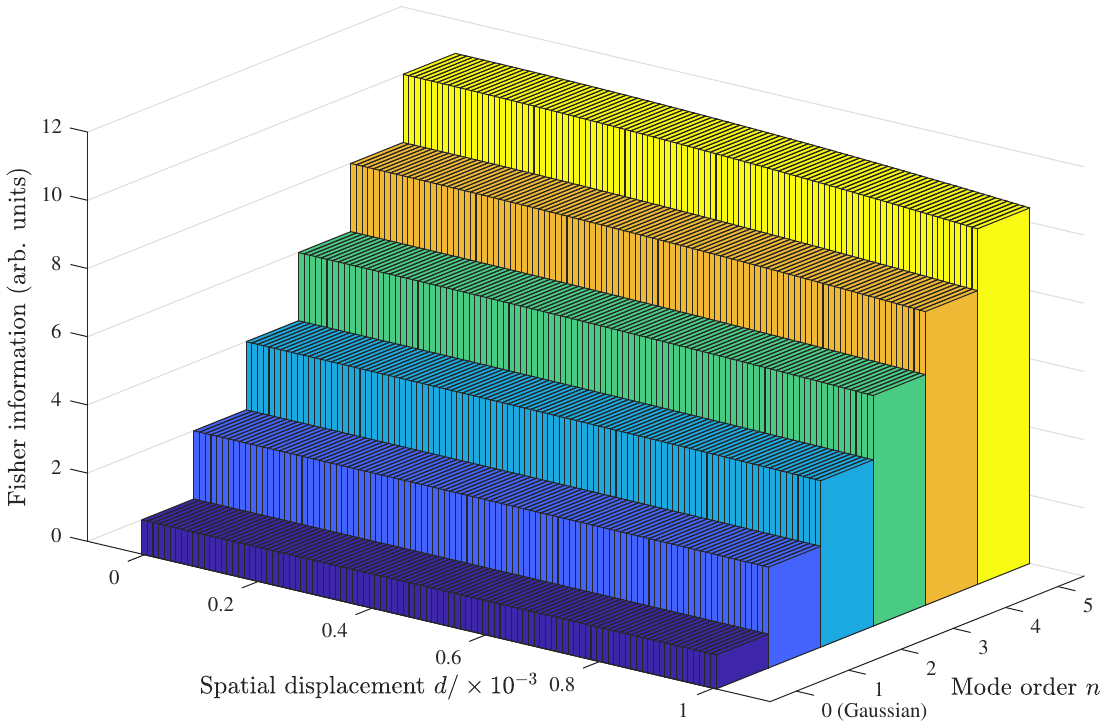}
			\label{fig:2-2a}
	\end{minipage}}
	\subfigure[Momentum kick]{
		\begin{minipage}{0.8\linewidth}
			\centering 
			\includegraphics[scale=0.37]{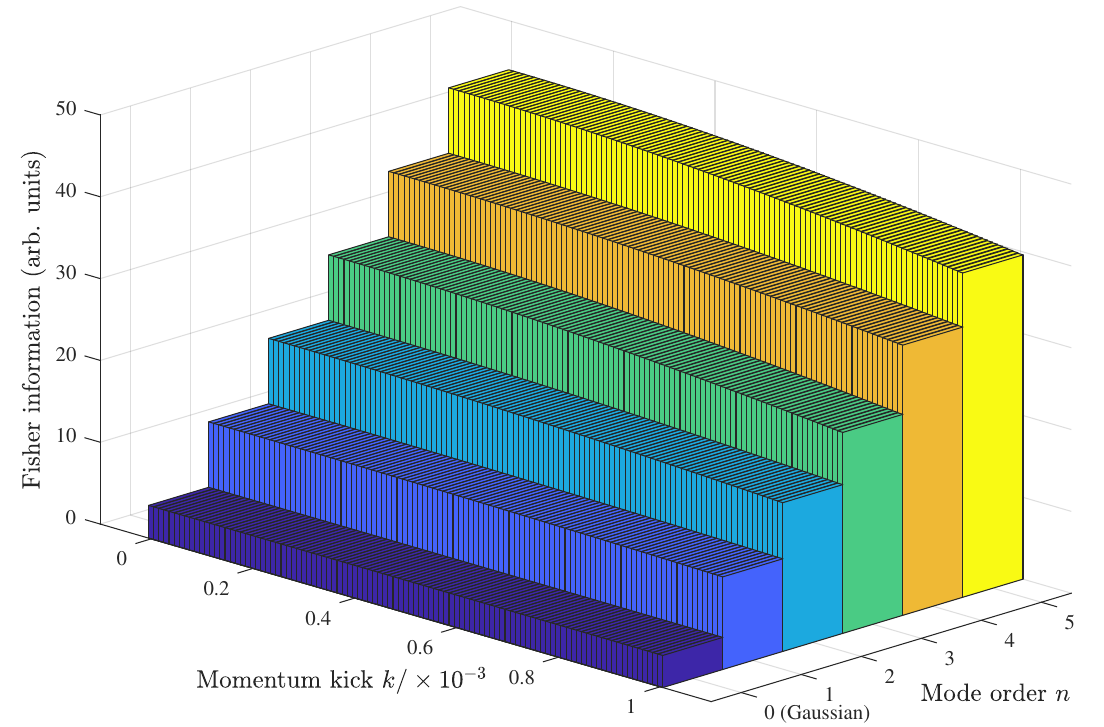}
			\label{fig:2-2b}
	\end{minipage}}
	\caption{Fisher information of spatial displacement and momentum kick. (a) Spatial displacement $d$ varies from 0 to 0.001, momentum kick $k=0$, Fisher information of spatial displacement $d$ from mode order 0 to 5. (b) Spatial displacement $d=0$, momentum kick $k$ varies from 0 to 0.001, Fisher information of momentum kick $k$ from mode order 0 to 5}
	\label{fig:2-2}
\end{figure}
The numerical simulation results are given in Fig.\ref{fig:2-2}. In Fig.\ref{fig:2-2a}, we chose $k=0$, and let $d$ varies from 0 to 0.001, the Fisher information of spatial displacement almost linearly increase by mode order $n$. Similarly, we chose $d=0$ and let $k$ varies from 0 to 0.001 in Fig.\ref{fig:2-2b}, the Fisher information of momentum kick also linearly increase by mode order $n$. This numerical results verified that our approximation result of QFIM is valid when $d\ll 1$ and $k\ll 1$.

We have proved the ultimate precision is improved by employing high-order H-G pointer. But in a practical measurement process, the final precision is restricted by the measurement strategy for final pointer state. A traditional measurement strategies is investigating the average position bias of pointer (light beam)\cite{RevModPhys.86.307,Turek_2015,PhysRevLett.109.230402}. Assuming $k=0$ in Eq.(\ref{eq:2-5}), the final position bias is
\begin{equation}
	\Delta x_{d} = \langle \phi_{f}|\hat{x}|\phi_{f}\rangle = \mathrm{Re}A_{w} d \nonumber
\end{equation}
Similarly, Assuming $d=0$ in Eq.(\ref{eq:2-5}), the final position bias is
\begin{equation}
	\Delta x_{k} = \langle \phi_{f}|\hat{x}|\phi_{f}\rangle = 2(2n+1)\sigma^{2}\mathrm{Im}A_{w} k \nonumber
\end{equation}
When independently measuring the spatial displacement, the result of amplification is unrelated to the mode order. But the amplification times increase with mode order when independently measuring the momentum kick. However, this result dose not mean average position measurement is available in Hermite-Gaussian mode situation\cite{Turek_2015}. The final position variance
\begin{equation}
	\Delta x_{f}^{2}\approx\Delta x_{i}^{2}=\langle\phi_{i}|\hat{x}^{2}|\phi_{i}\rangle=(2n+1)\sigma^{2} \nonumber
\end{equation}
also increase with mode order $n$. That means the measurement SNR of Hermite-Gaussian pointer is not better than Gaussian pointer.

Although traditional average position bias measurement is failed to improve precision with H-G mode pointer, the application of H-G pointer is not nonsense. We have proposed two different measurement strategies, maximum likelihood estimation (MLE) and homodyne detection, which can approximately approach the QFIM limit.
\section{Measurement Strategies} \label{sec:3}
In this part, we introduce two kind of measurement strategies about final pointer state $|\phi_{f}\rangle$. One method is measuring the spatial distribution of $|\phi_{f}\rangle$, and then applying maximum likelihood estimation (MLE)\cite{2018arXiv180604503V} to this final spatial distribution. This method is widely used in some previous researches about one parameter weak measurement\cite{PhysRevX.4.011031,Fang_2016,2019arXiv190106831M,PhysRevA.96.032112}. The other one is homodyne detection\cite{Liu:17,PhysRevA.74.053823,doi:10.1063/1.4869819}, which uses a neighbour mode local-oscillator light beam to interfere with the measurement beam. We note that the main information about $d$ and $k$ is carried by mode $n-1$ and $n+1$. Thus we can employ homodyne detection to extract information about $d$ and $k$ in mode $n-1$ and $n+1$ of final pointer state $|\phi_{f}\rangle$.

In last section, we gave the precision limit of weak measurement about $\mathbf{g}$ with Hermite-Gaussian pointer from quantum Fisher information, which is the measurement (of the final pointer state) independent result. To calculate the ultimate precision a certain detection strategy can achieve, classical Fisher information matrix (CFIM) is employed. Similarly, the inverse of CFIM, classical Cram$\mathrm{\acute{e}}$r-Rao bound (CCRB), gives the low bound of estimation covariance matrix about parameter vector $\mathbf{g}$ with a certain measurement strategy of final pointer state $|\phi_{f}\rangle$.
\subsection{Maximum Likelihood Estimation on Pointer Distribution}
First, the final pointer spatial distribution is given by:
\begin{equation}
	P_{f}(x|\mathbf{g}) = |\langle x|\phi_{f}\rangle|^{2} \label{eq:3-1}
\end{equation}
In the case of performing MLE on the pointer distribution, the corresponding CFIM can be calculated by:
\begin{equation}
	[\mathbb{F}^{(M)}]_{ij} = \int \mathrm{d}x\frac{1}{P_{f}(x|\mathbf{g})}\left[\frac{\partial P_{f}(x|\mathbf{g})}{\partial g_{i}}\frac{\partial P_{f}(x|\mathbf{g})}{\partial g_{j}}\right] \label{eq:3-2}
\end{equation}
which is the $i-$th row, $j-$th column component of matrix $\mathbb{F}^{(M)}$. Plugging Eq.(\ref{eq:2-5}) into Eq.(\ref{eq:3-1}), then calculating every components in matrix $\mathbb{F}^{(M)}$ by Eq.(\ref{eq:3-2}). Based on the calculation in App.\ref{app:1}, we finally get the CFIM of MLE strategy:
\begin{widetext}
\begin{equation}
\label{eq:3-3}
	\mathbb{F}^{(M)}_{N'} = |\langle f|i\rangle|^{2}N\left(
	\begin{array}{cc}
	(2n+1)(\mathrm{Re}A_{w})^{2}\sigma^{-2} & 2(\mathrm{Re}A_{w})(\mathrm{Im}A_{w})\\
	2(\mathrm{Re}A_{w})(\mathrm{Im}A_{w}) & 4(2n+1)(\mathrm{Im}A_{w})^{2}\sigma^{2}
	\end{array}
	\right)
\end{equation}	
\end{widetext}
Here we give the $N'$ time result of CFIM in the MLE case directly. As is stated above, the inverse of $\mathbb{F}^{(M)}_{N'}$ gives a low bound precision for estimation of parameter $\mathbf{g}$ with MLE method, that we can describe it as $\mathbb{C}_{N'}^{(M)}\ge[\mathbb{F}_{N'}^{(M)}]^{-1}$. Calculating the inverse of $\mathbb{F}^{(M)}_{N'}$, we find that the covariance of $d$ and $k$ is not 0. For Gaussian mode pointer ($n=0$), $|\mathbb{F}_{N'}^{(M)}|=0$, $[\mathbb{F}_{N'}^{(M)}]^{-1}$ is an infinite value matrix. It means that the estimation precision of Gaussian pointer in MLE case is nearly infinite, Gaussian pointer is not properly for our multi-parameter estimation in MLE case anymore.

\begin{widetext}
\begin{figure*}[ht]
	\centering
	\subfigure[n=1]{
		\begin{minipage}{0.3\linewidth}
			\centering  
			\includegraphics[scale=0.3]{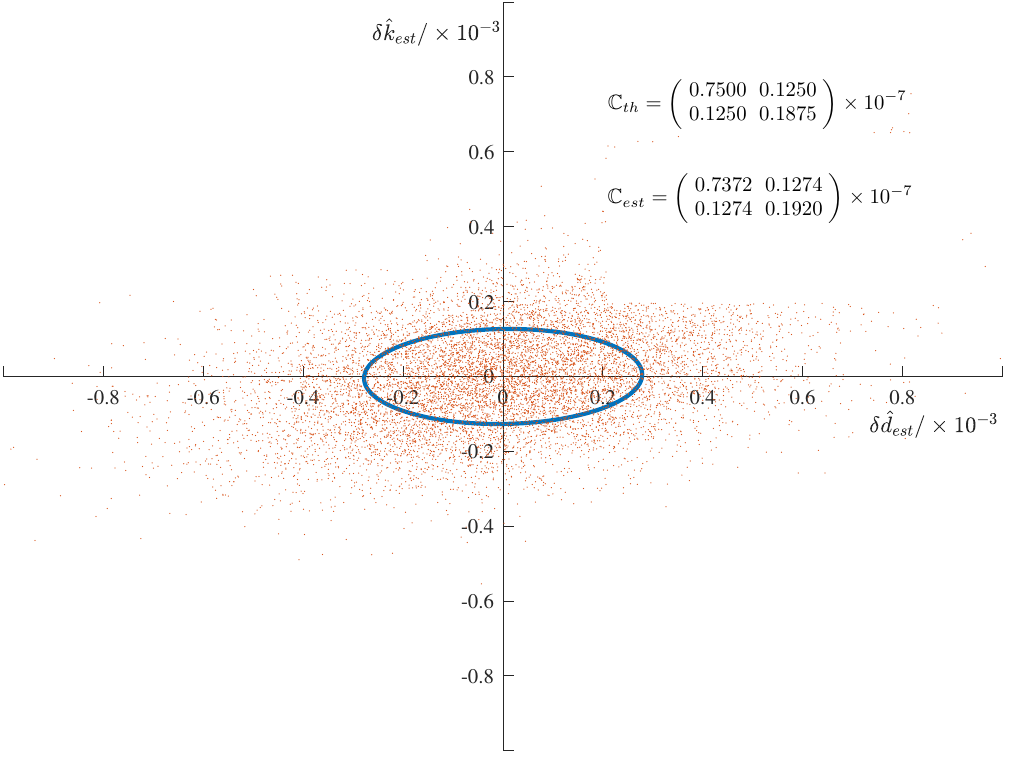}
			\label{fig:3-1a}
	\end{minipage}}
	\subfigure[n=2]{
		\begin{minipage}{0.3\linewidth}
			\centering 
			\includegraphics[scale=0.3]{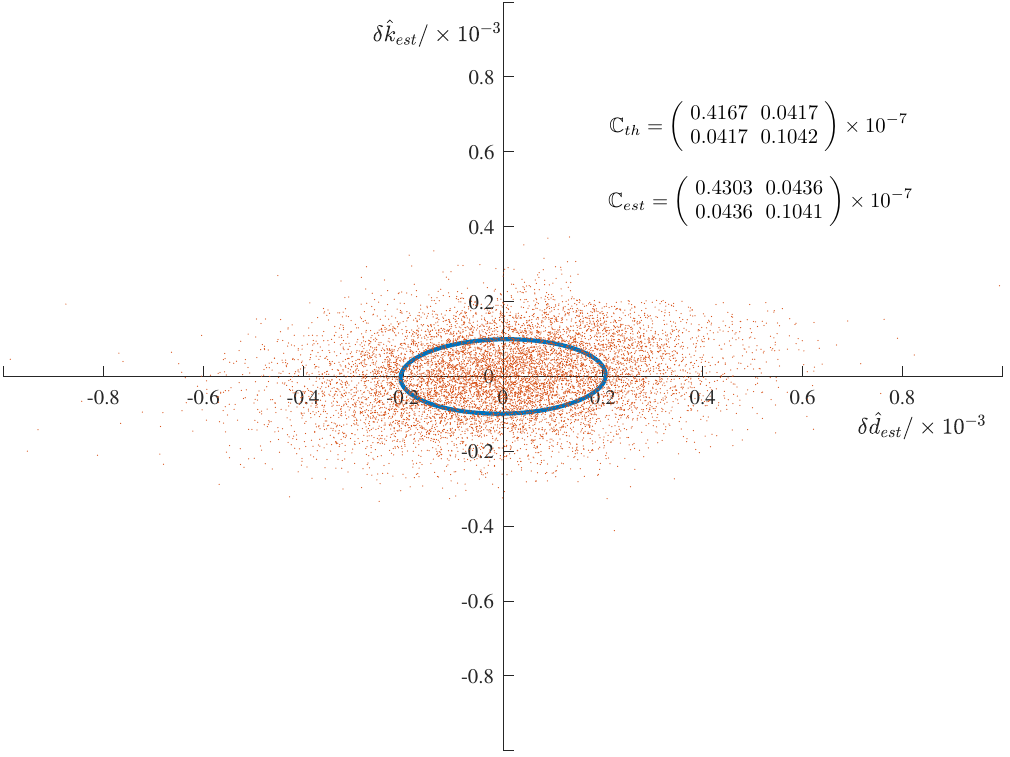}
			\label{fig:3-1b}
	\end{minipage}}
	\subfigure[n=3]{
		\begin{minipage}{0.3\linewidth}
			\centering 
			\includegraphics[scale=0.3]{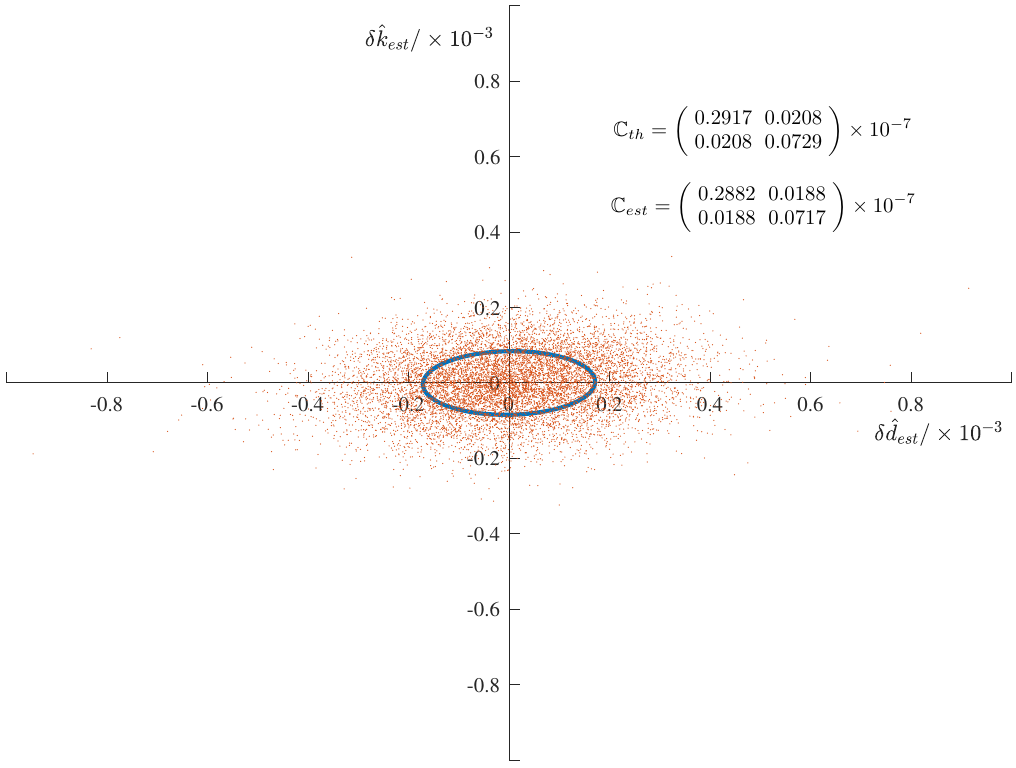}
			\label{fig:3-1c}
	\end{minipage}}
	\subfigure[n=4]{
		\begin{minipage}{0.4\linewidth}
			\centering 
			\includegraphics[scale=0.3]{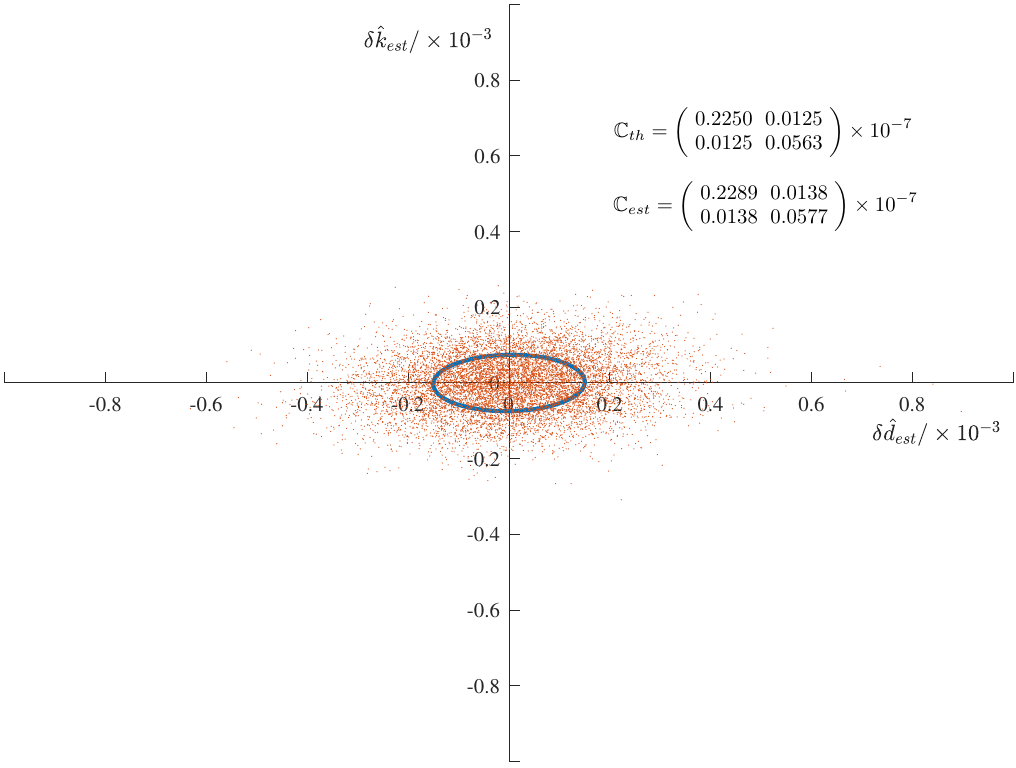}
			\label{fig:3-1d}
	\end{minipage}}
	\subfigure[n=5]{
		\begin{minipage}{0.4\linewidth}
			\centering 
			\includegraphics[scale=0.3]{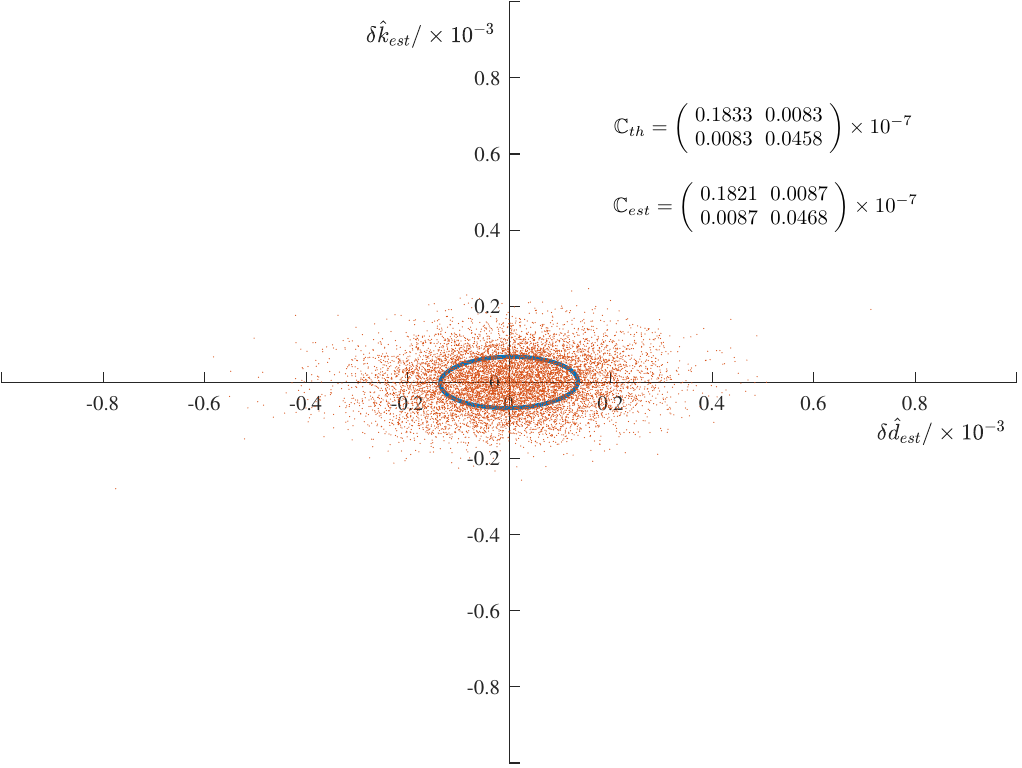}
			\label{fig:3-1e}
	\end{minipage}}
	\caption{Error eclipse. Blue eclipse: error eclipse calculated by theoretical covariance matrix. Orange dot: Simulation results (difference value between $(\hat{d}_{est},\hat{k}_{est})$ and pre-set value (truth-value) of $(d,k)$). $\mathbb{C}_{th}$ is the theoretical covariance matrix, $\mathbb{C}_{est}$ is the simulation covariance matrix.}
	\label{fig:3-1}
\end{figure*}
\end{widetext}

\paragraph*{Tradeoff} From Eq.(\ref{eq:2-9}) and Eq.(\ref{eq:3-3}), we can derive a tradeoff relation between QFIM and CFIM, $\mathrm{Tr}[\mathbb{F}_{N'}^{(M)}(\mathbb{Q}_{N'})^{-1}]=1$. In CFIM, the Fisher  information about spatial displacement $d$ comes from the real part of weak value $A_{w}$, and the Fisher  information about momentum kick $k$ comes from the image part of weak value $A_{w}$. That means if weak value $A_{w}$ is a real value, the single measurement of $d$ can saturate the QFI limit with MLE method, and if $A_{w}$ is a pure image value, the single measurement of $k$ can saturate the QFI limit. But these two QFI limits couldn't be saturated simultaneously because of the tradeoff relation $\mathrm{Tr}[\mathbb{F}_{N'}^{(M)}(\mathbb{Q}_{N'})^{-1}]=1$.
\paragraph*{Simulation} In this part, we give a simulation with MLE method for measurement of $d$ and $k$. Supposed that the detection result is a group of position sample ${x_{i}}$, thus the log-likelihood function can be given as $\mathcal{L}(\mathbf{g}|{x_{i}})=\sum_{i}P_{f}(x_{i}|\mathbf{g})$. Maximize the log-likelihood function, we can solve out the parameter vector $\mathbf{g}$, which is the estimation results of $\hat{d}_{est}$ and $\hat{k}_{est}$. Here, we still choose pre-selection $|i\rangle=\frac{1}{\sqrt{2}}(|H\rangle+|V\rangle)$ and post-selection $|f\rangle=e^{i\frac{\varepsilon}{2}}\cos (\frac{\pi}{4}+\frac{\varepsilon}{2})|H\rangle-e^{-i\frac{\varepsilon}{2}}\sin (\frac{\pi}{4}+\frac{\varepsilon}{2})|V\rangle$, where $\varepsilon=0.01$. Operator $\hat{A}=|H\rangle\langle H|-|V\rangle\langle V|$ is Pauli operator in this simulation. And the parameter $\sigma$ in pointer's transverse spatial distribution is set to 1. We using Eq.(\ref{eq:3-1}) to generate groups of random dots, which can be explained as the detection results. We suppose the detected sample number is $N'=500$, which means the source sample number $N=N'/|\langle f|i\rangle|^{2}\approx 10^{7}$. In Fig.\ref{fig:3-1}, we pre-set $d=0$ and $k=0$, then generate 10000 groups simulation dots for mode order 1 to 5, each group contains 500 samples. Then we employ the MLE method on these samples to calculate the $\hat{d}_{est}$ and $\hat{k}_{est}$, the error eclipses from mode order 1 to 5 are given in Fig.\ref{fig:3-1}.
\subsection{Homodyne Detection} \label{sec:3-2}
First, we introduce a general local-oscillator state $|\phi^{LO}\rangle=\alpha|\phi_{n-1}\rangle+\beta|\phi_{n+1}\rangle$, where $|\alpha|^{2}+|\beta|^{2}=1$. The final difference intensity of homodyne will give information about parameter $\mathbf{g}$. But here we have two unknown parameters $d$ and $k$, one homodyne detection can only give one useful information, two independent homodyne detection is necessary in our multi-parameter estimation scenario. 
\begin{figure}[h]
	\centering
	\includegraphics[scale=0.5]{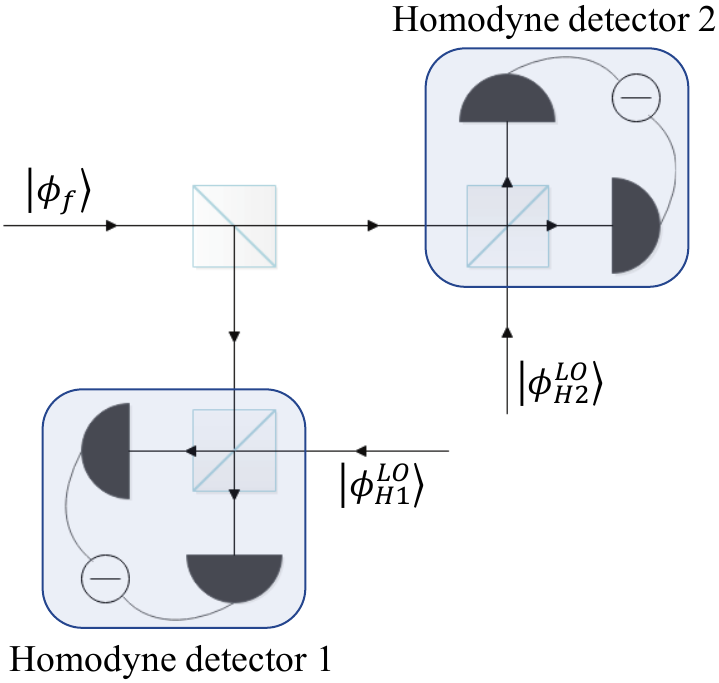}
	\caption{\label{fig:3-3}Homodyne detection scheme. The final pointer are divided into two part by a 50:50 beam splitter for two independent homodyne detection.}
\end{figure}

The scheme of homodyne detection is shown in Fig.\ref{fig:3-3}, these two different local-oscillator state is
\begin{subequations}
	\label{eq:3-4}
	\begin{equation}
		|\phi^{LO}_{H1}\rangle=\alpha_{1}|\phi_{n-1}\rangle+\beta_{1}|\phi_{n+1}\rangle \label{eq:3-4a}
	\end{equation}
	\begin{equation}
		|\phi^{LO}_{H2}\rangle=\alpha_{2}|\phi_{n-1}\rangle+\beta_{2}|\phi_{n+1}\rangle \label{eq:3-4b}
	\end{equation}
\end{subequations}
where $|\alpha_{i}|^{2}+|\beta_{i}|^{2}=1$, $i=1,2$. Because the pointer and local-oscillator state may have different amplitude strengths, we need take this into consideration. Thus these two received states in detector 1 is denoted as:
\begin{subequations}
	\label{eq:sup3-1}
	\begin{equation}
	|\phi_{H1}^{+}\rangle=\frac{1}{\sqrt{2}}(A_{f1}|\phi_{f}\rangle+A_{LO1}|\phi^{LO}_{H1}\rangle) \label{eq:sup3-1a}
	\end{equation}
	\begin{equation}
	|\phi_{H1}^{-}\rangle=\frac{1}{\sqrt{2}}(A_{f1}|\phi_{f}\rangle-A_{LO1}|\phi^{LO}_{H1}\rangle) \label{eq:sup3-1b}
	\end{equation}
\end{subequations}
Similarly, the received states in detector 2 is:
\begin{subequations}
	\label{eq:sup3-2}
	\begin{equation}
	|\phi_{H2}^{+}\rangle=\frac{1}{\sqrt{2}}(A_{f2}|\phi_{f}\rangle+A_{LO2}|\phi^{LO}_{H2}\rangle) \label{eq:sup3-2a}
	\end{equation}
	\begin{equation}
	|\phi_{H2}^{-}\rangle=\frac{1}{\sqrt{2}}(A_{f2}|\phi_{f}\rangle-A_{LO2}|\phi^{LO}_{H2}\rangle) \label{eq:sup3-2b}
	\end{equation}
\end{subequations}
where $A_{f1}$ and $A_{f2}$ are corresponding amplitude strength of final pointer in detector 1 and 2, $A_{LO1}$ and $A_{LO2}$ are corresponding amplitude strength of local oscillator in detector 1 and 2.

Thus the detected intensities of these detection ports in homodyne detector 1 and detector 2 are:
\begin{gather}
	I_{H1}^{+}=\langle\phi_{H1}^{+}|\phi_{H1}^{+}\rangle;\quad I_{H1}^{-}=\langle\phi_{H1}^{-}|\phi_{H1}^{-}\rangle \label{eq:3-5}\\
	I_{H2}^{+}=\langle\phi_{H2}^{+}|\phi_{H2}^{+}\rangle;\quad I_{H2}^{-}=\langle\phi_{H2}^{-}|\phi_{H2}^{-}\rangle \label{eq:3-6}
\end{gather}
Suppose that the input intensity of two homodyne detector is same, the relations between amplitude strength and photons' number is $A_{f}=\sqrt{2}A_{f1}=\sqrt{2}A_{f2}=\sqrt{N'}$, $A_{LO}=\sqrt{2}A_{LO1}=\sqrt{2}A_{LO2}=\sqrt{N_{LO}}$, where $N'=|\langle f|i\rangle|^{2}N$ is the number of measurement photons, and $N_{LO}$ is the number of local oscillator.

The corresponding probabilities of every detection port is
\begin{equation}
	\label{eq:3-7}
	\begin{split}
	P_{H1}^{+}=\frac{I_{H1}^{+}}{\sum_{i} I_{Hi}^{+}+I_{Hi}^{-}},\; P_{H1}^{-}=\frac{I_{H1}^{-}}{\sum_{i} I_{Hi}^{+}+I_{Hi}^{-}}\\
	P_{H2}^{+}=\frac{I_{H2}^{+}}{\sum_{i} I_{Hi}^{+}+I_{Hi}^{-}},\; P_{H2}^{-}=\frac{I_{H2}^{-}}{\sum_{i} I_{Hi}^{+}+I_{Hi}^{-}}
	\end{split}
\end{equation}
Thus the CFIM of homodyne detection can be calculated by
\begin{eqnarray}
	[\mathbb{F}_{N'}^{(H)}]_{ij} =&& \sum_{m=1,2} \frac{1}{P_{Hm}^{+}}\left[\frac{\partial P_{Hm}^{+}}{\partial g_{i}}\frac{\partial P_{Hm}^{+}}{\partial g_{j}}\right] \nonumber\\
	&&+\sum_{m=1,2}\frac{1}{P_{Hm}^{-}}\left[\frac{\partial P_{Hm}^{-}}{\partial g_{i}}\frac{\partial P_{Hm}^{-}}{\partial g_{j}}\right] \label{eq:3-8}
\end{eqnarray}
where $g_{1}=d$, $g_{2}=k$.

In App.\ref{app:2}, we have proved that the tradeoff relation $\mathrm{Tr}[\mathbb{F}_{N'}^{(H)}(\mathbb{Q}_{N'})^{-1}]\le 1$ always holds. It means that the quantum precision limits of parameters $d$ and $k$ can't be saturated at same time either by homodyne detection method. But this method is easier to implement in experiment than MLE method, and Gaussian mode pointer is not useless in this case.

The maximum value of $\mathrm{Tr}[\mathbb{F}_{N'}^{(H)}(\mathbb{Q}_{N'})^{-1}]$ is 1. We derived the conditions for this maximum value in App.\ref{app:2}, which is $\mathrm{Re}(A_{w}^{2}\alpha_{i}^{*}\beta_{i}^{*})=-|A_{w}\alpha_{i}^{*}|\cdot|A_{w}\beta_{i}^{*}|$ and $|\alpha_{i}|/|\beta_{i}|=\sqrt{n}/\sqrt{n+1}$, where $i=1,2$. Here we give a typical example of local oscillator for instance:
\begin{subequations}
	\label{eq:3-9}
	\begin{equation}
		|\phi^{LO}_{H1}\rangle=\frac{1}{\sqrt{2n+1}}(\sqrt{n}|\phi_{n-1}\rangle+i\sqrt{n+1}|\phi_{n+1}\rangle) \label{eq:3-9a}
	\end{equation}
	\begin{equation}
		|\phi^{LO}_{H2}\rangle=\frac{1}{\sqrt{2n+1}}(i\sqrt{n}|\phi_{n-1}\rangle+\sqrt{n+1}|\phi_{n+1}\rangle) \label{eq:3-9b}
	\end{equation}
\end{subequations}
Combining Eq.(\ref{eq:3-8}) with Eq.(\ref{eq:sup3-1}) to Eq.(\ref{eq:3-8}), the CFIM of homodyne detection method is given as:
\begin{equation}
	\label{eq:3-10}
	\mathbb{F}_{N'}^{(H)} = |\langle f|i\rangle|^{2}N(2n+1)|A_{w}|^{2}\left(
	\begin{array}{cc}
	\frac{1}{2}\sigma^{-2} & 0\\
	0 & 2\sigma^{2}
	\end{array}
\right)
\end{equation}
Here we use hypotheses $N_{LO}\gg N'=|\langle f|i\rangle|^{2}N$, and choose $\mathrm{Re}A_{w}=-\mathrm{Im}A_{w}$ to get maximum value of $-\mathrm{Re}A_{w}\mathrm{Im}A_{w}$, which is $\frac{1}{2}|A_{w}|^{2}$. The calculation details are given in supplemental materials.

Although MLE and homodyne method can approach the ultimate precision in estimating single parameter, neither MLE method nor homodyne detection can saturate QFIM when simultaneously measuring every parameters. The trace $\mathrm{Tr}[\mathbb{F}_{N'}(\mathbb{Q}_{N'})^{-1}]$ is not more than 1 with these two measurement strategies. But we have found that $\langle\phi_{f}|L_{d}L{k}-L_{k}L_{d}|\phi_{f}\rangle=0$, which implies the QFIM can be saturated when simultaneously measuring every parameters\cite{Matsumoto_2002,2018arXiv180607337Y}. Though we haven't known what the optimal POVM formalism is, this result is helpful for our successive research.
\section{Experimental Setup} \label{sec:4}
\begin{figure}[h]
	\centering
	\includegraphics[scale=0.3]{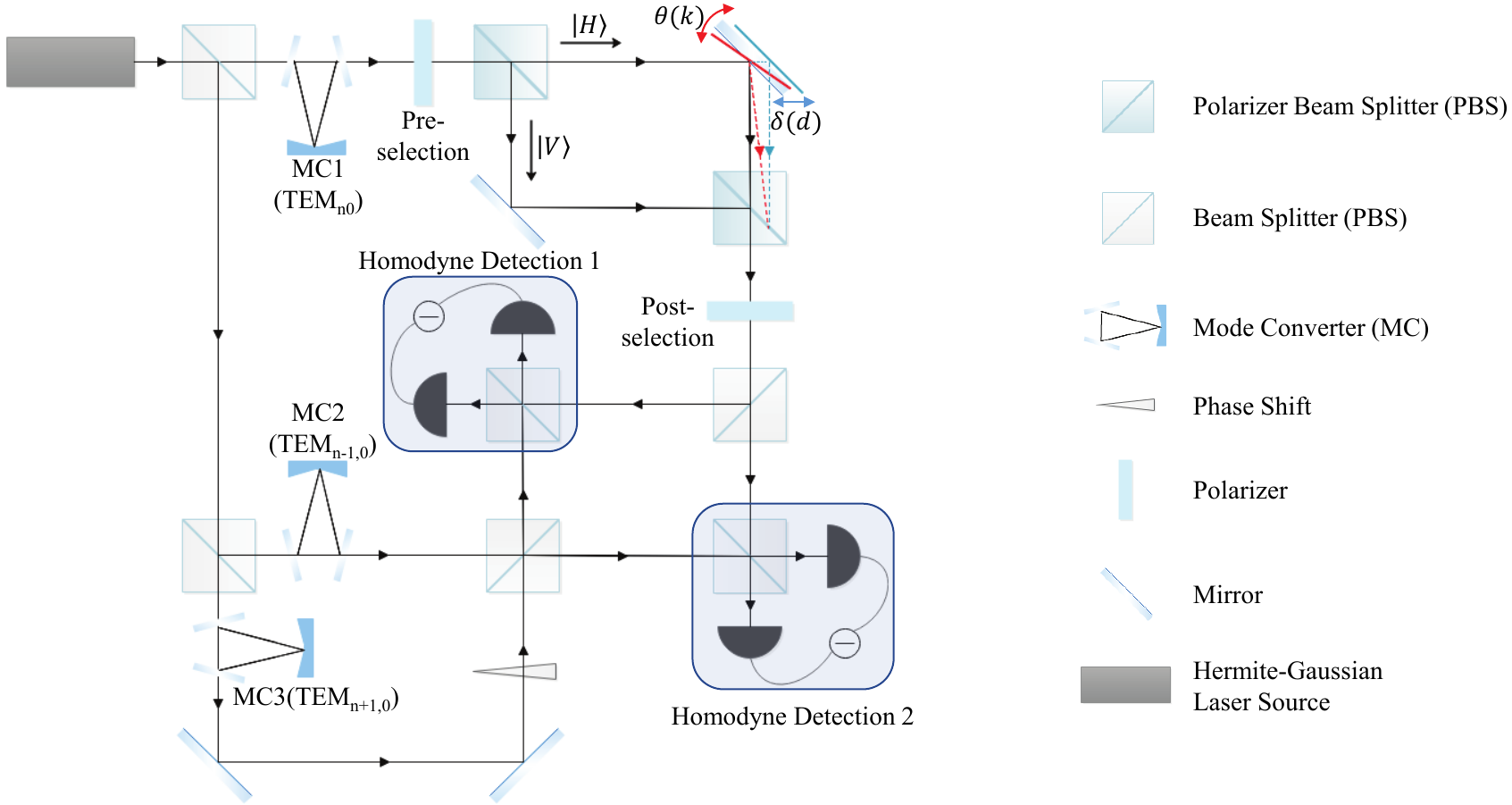}
	\caption{\label{fig:4-1}The schematic paradigm.}
\end{figure}

In the practical optical experiment, the Hermite-Gaussian mode is usually carried by the transverse distribution of light beam. Thus the spatial displacement $d$ is a little transverse displacement of light beam and momentum kick $k$ is a little transverse kick. Combining the results derived in Sec.\ref{sec:3-2}, we propose a proof-of-principle weak measurement scheme in this part, where the pointer is a light beam with $n$-th order H-G transverse distribution, $d$ and $k$ are caused by the mirror's displacement and tilt separately. The experimental setup is depicted in Fig.\ref{fig:4-1}.

In this scheme, the import light beam is Hermite-Gaussian mode, MC1 modulate the measurement beam (pointer) in $TEM_{n,0}$ mode. MC2 and MC3 are employed to modulate local-oscillator beam in $TEM_{n-1,0}$ and $TEM_{n+1,0}$. The weak interaction process is realized by a Mach–Zehnder interferometer as shown in Fig.\ref{fig:4-1}.Thus the measurement operator $\hat{A}=\hat{\sigma}_{z}+\hat{\mathbb{I}}$, where $\hat{\sigma}_{z}=|H\rangle\langle H|-|V\rangle\langle V|$ is Pauli operator. The pre-selected and post-selected polarizing states are $|i\rangle=\frac{1}{\sqrt{2}}(|H\rangle+|V\rangle)$ and $|f\rangle=e^{i\frac{\varepsilon}{2}}\cos(\frac{\pi}{4}+\frac{\varepsilon}{2})|H\rangle-e^{-i\frac{\varepsilon}{2}}\sin(\frac{\pi}{4}+\frac{\varepsilon}{2})|V\rangle$, where $\varepsilon\ll 1$.Thus the weak value in our experimental scheme is given as:
\begin{equation}
	A_{w}=\frac{\langle f|\hat{A}|i\rangle}{\langle f|i\rangle}\approx -\frac{1}{\varepsilon}+i\frac{1}{\varepsilon} \label{eq:4-1}
\end{equation}

We inject two light beams with $TEM_{n-1,0}$ and $TEM_{n+1,0}$ mode into a 50:50 beam splitter, these two emergent beam are the local-oscillator lights we used. Thus the amplitude function of local-oscillator lights are:
\begin{subequations}
	\label{eq:4-2}
	\begin{equation}
		\phi^{LO}_{H1}(x)=\frac{1}{\sqrt{2}}[\phi_{n-1}(x)+i\phi_{n+1}(x)] \label{eq:4-2a}
	\end{equation}
	\begin{equation}
		\phi^{LO}_{H2}(x)=\frac{1}{\sqrt{2}}[i\phi_{n-1}(x)+\phi_{n+1}(x)] \label{eq:4-2b}
	\end{equation}
\end{subequations}
Combining Eq.(\ref{eq:4-2}) with Eq.(\ref{eq:sup3-1}) to Eq.(\ref{eq:3-6}), we have the difference intensity of homodyne detection 1 is:
\begin{align}
	\Delta I_{H1} &= I_{H1}^{+}-I_{H1}^{-} \nonumber\\
	 &= \frac{1}{\sqrt{2}\varepsilon}A_{f}A_{LO}(\sqrt{n}+\sqrt{n+1})(\frac{1}{2\sigma}d+\sigma k) \label{eq:4-3}
\end{align}
Similarly, the difference intensity of homodyne detection 2 is:
\begin{align}
	\Delta I_{H2} &= I_{H2}^{+}-I_{H2}^{-} \nonumber\\
	&= \frac{1}{\sqrt{2}\varepsilon}A_{f}A_{LO}(\sqrt{n}+\sqrt{n+1})(-\frac{1}{2\sigma}d+\sigma k) \label{eq:4-4}
\end{align}
Combining Eq.(\ref{eq:4-3}) and Eq.(\ref{eq:4-4}), the spatial displacement and momentum kick of light are separately derived as:
\begin{subequations}
	\label{eq:4-5}
	\begin{equation}
		\hat{d}_{est}=\frac{\sqrt{2}\varepsilon\sigma}{\sqrt{n}+\sqrt{n+1}}\frac{\Delta I_{H1}-\Delta I_{H2}}{A_{f}A_{LO}} \label{eq:4-5a}
	\end{equation}
	\begin{equation}
		\hat{k}_{est}=\frac{\varepsilon}{\sqrt{2}\sigma}\frac{1}{\sqrt{n}+\sqrt{n+1}}\frac{\Delta I_{H1}+\Delta I_{H2}}{A_{f}A_{LO}} \label{eq:4-5b}
	\end{equation}
\end{subequations}
where $A_{f}=\sqrt{N'}$ and $A_{LO}=\sqrt{N_{LO}}$. The displacement of mirror $\delta=\hat{d}_{est}/2$, the angular tilt of mirror $\theta=\frac{\lambda_{0}}{8\pi}\hat{k}_{est}$, where $\lambda_{0}$ is the wave length of source.

In classical theory, the precision limit of measurement is determined by the shot noise, which is proportional to $\sqrt{N'+N_{LO}}$. In the shot noise limit, the minimal detectable mirror displacement is
\begin{eqnarray}
	\delta_{min} &&= \frac{1}{2}d_{min} = \frac{1}{2}\frac{\sqrt{2}\varepsilon\sigma}{\sqrt{n}+\sqrt{n+1}}\sqrt{\frac{1}{N_{LO}}+\frac{2}{\varepsilon^{2}N}} \nonumber\\
	&&\approx \frac{\sigma}{\sqrt{n}+\sqrt{n+1}}\frac{1}{\sqrt{N}} \label{eq:4-6}
\end{eqnarray}
and the minimal detectable mirror angular tilt is
\begin{eqnarray}
	\theta_{min} &&= \frac{\lambda_{0}}{8\pi}k_{min} = \frac{\lambda_{0}\varepsilon}{8\sqrt{2}\pi\sigma(\sqrt{n}+\sqrt{n+1})}\sqrt{\frac{1}{N_{LO}}+\frac{2}{\varepsilon^{2}N}} \nonumber\\
	&&\approx \frac{\lambda_{0}}{8\pi\sigma(\sqrt{n}+\sqrt{n+1})}\frac{1}{\sqrt{N}} \label{eq:4-7}
\end{eqnarray}
The ultimate precision in our experimental scheme increases with spatial mode order $n$ of source light by factor $\frac{1}{\sqrt{2}}(\sqrt{n}+\sqrt{n+1})$. In Sec.\ref{sec:3-2}, we derived an optimal increase factor is $\sqrt{2n+1}$. But the LO light form in Sec.\ref{sec:3-2} is not easy to prepare. Our scheme uses a beam splitter to produce two LO lights, which losses precision but is much easier to implement.

Finally, we note that in our theoretical analysis and experimental setup, the H-G modes employed for pointer and local-oscillator are assumed to be pure. However in practice, the generation process of light could not be perfect, thus further investigations on the effects of H-G mode fidelity are required.

\section{Discussions and Conclusions} \label{sec:5} 
In summary, we have established a general multi-parameter weak measurement formalism, and investigated a common weak measurement scenario with unknown spatial displacement $d$ and momentum kick $k$. To improve the measurement precision, high-order H-G pointers are employed in our weak measurement scheme. Theoretical analysis based on QFIM shows that the ultimate precision linearly increases with H-G mode order by a factor of $\sqrt{2n+1}$. Moreover, we proposed two available measurement strategies, i.e. MLE and homodyne detection. Analysis based on CFIM reveal that both MLE and homodyne methods can approach the ultimate precision limit given by QFIM. Finally, we have proposed a proof-of-principle experimental scheme simultaneously measuring ultra-small object displacement and tilt with $n$-th order H-G mode laser and optical homodyne detection. In principle, the experimental can be improved hy a factor of $\frac{1}{\sqrt{2}}(\sqrt{n}+\sqrt{n+1})$ comparing to the conventional WVA scheme with fundamental Gaussian pointer. 
Moreover, our results can be extended by expressing the pointer in other spacial-mode basis, such as the Laguerre-Gaussian (L-G) modes. As the quantum counterpart of L-G mode, i.e. the quantum optical angular momentum (OAM)\cite{Padgett:17}, was successfully adopted as pointer in the weak measurement experiment\cite{PhysRevLett.112.200401}, it would be interesting to generalize our theoretical framework for larger range of applications. 

\begin{acknowledgments}
	This work was supported by the National Natural Science Foundation of China (Grants Nos. 61671287, 61631014, 61701302, 61901258), and the fund of State Key Laboratory of Advanced Optical Communication Systems and Networks.
\end{acknowledgments}

\begin{widetext}
\appendix

\section{CFIM of MLE Strategy} \label{app:1}
Plugging Eq.(\ref{eq:sup2-4}) into Eq.(\ref{eq:3-1}), we have:
\begin{equation}
	\label{eq:a1-1}
	P_{f}(x|\mathbf{g})=\phi_{n}^{2}(x)-2(\mathrm{Re}A_{w})(\partial_{x}\phi_{n}(x))\phi_{n}(x)d+2(\mathrm{Im}A_{w})x\phi_{n}^{2}(x)k
\end{equation}
where $\mathbf{g}=(d,k)$ is the unknown parameter vector, and we only take first order result about $\mathbf{g}$ because of $d\ll 1$ and $k\ll 1$. The partial differential results of $P_{f}(x|\mathbf{g})$ is calculated as:
\begin{subequations}
	\label{eq:a1-2}
	\begin{equation}
		\frac{\partial P_{f}(x|\mathbf{g})}{\partial d}=-2(\mathrm{Re}A_{w})(\partial_{x}\phi_{n}(x))\phi_{n}(x) \label{eq:a1-2a}
	\end{equation}
	\begin{equation}
		\frac{\partial P_{f}(x|\mathbf{g})}{\partial k}=2(\mathrm{Im}A_{w})x\phi_{n}^{2}(x) \label{eq:a1-2b}
	\end{equation}
\end{subequations}
Combining Eq.(\ref{eq:a1-2}) with Eq.(\ref{eq:3-2}), we can separately calculate the Fisher information about $d$:
\begin{align}
	[\mathbb{F}^{(M)}]_{11} =& \int \mathrm{d}x\frac{1}{P_{f}(x|\mathbf{g})}\left[\frac{\partial P_{f}(x|\mathbf{g})}{\partial d}\right]^{2} \nonumber\\
	\approx& \int \mathrm{d}x\, 4(\mathrm{Re}A_{w})^{2}(\partial_{x}\phi_{n}(x))^{2}\left[1+2(\mathrm{Re}A_{w})\frac{\partial_{x}\phi_{n}(x)}{\phi_{n}(x)}d-2(\mathrm{Im}A_{w})xk\right] \nonumber\\
	\approx& 4(\mathrm{Re}A_{w})^{2}\int (\partial_{x}\phi_{n}(x))^{2}\mathrm{d}x \nonumber\\
	=& (2n+1)(\mathrm{Re}A_{w})^{2}\sigma^{-2} \label{eq:a1-3}
\end{align}
Here, we only care about the consistent term in Fisher information. Similarly, the result about $k$ is:
\begin{align}
	[\mathbb{F}^{(M)}]_{22} =& \int \mathrm{d}x\frac{1}{P_{f}(x|\mathbf{g})}\left[\frac{\partial P_{f}(x|\mathbf{g})}{\partial k}\right]^{2} \nonumber\\
	\approx& \int \mathrm{d}x\, 4(\mathrm{Im}A_{w})^{2}x^{2}\phi_{n}^{2}(x)\left[1+2(\mathrm{Re}A_{w})\frac{\partial_{x}\phi_{n}(x)}{\phi_{n}(x)}d-2(\mathrm{Im}A_{w})xk\right] \nonumber\\
	\approx& 4(\mathrm{Im}A_{w})^{2}\int x^{2}\phi_{n}(x)^{2}\mathrm{d}x \nonumber\\
	=& 4(2n+1)(\mathrm{Im}A_{w})^{2}\sigma^{2} \label{eq:a1-4}
\end{align}
And their correlation Fisher information is:
\begin{align}
	[\mathbb{F}^{(M)}]_{12} = [\mathbb{F}^{(M)}]_{21} =& \int \mathrm{d}x\frac{1}{P_{f}(x|\mathbf{g})}\left[\frac{\partial P_{f}(x|\mathbf{g})}{\partial d}\frac{\partial P_{f}(x|\mathbf{g})}{\partial k}\right] \nonumber\\
	\approx& \int \mathrm{d}x\, [-4(\mathrm{Re}A_{w})(\mathrm{Im}A_{w})x(\partial_{x}\phi_{n}(x))\phi_{n}(x)]\left[1+2(\mathrm{Re}A_{w})\frac{\partial_{x}\phi_{n}(x)}{\phi_{n}(x)}d-2(\mathrm{Im}A_{w})xk\right] \nonumber\\
	\approx& -4(\mathrm{Re}A_{w})(\mathrm{Im}A_{w})\int x(\partial_{x}\phi_{n}(x))\phi_{n}(x)\mathrm{d}x \nonumber\\
	=& 2(\mathrm{Re}A_{w})(\mathrm{Im}A_{w}) \label{eq:a1-5}
\end{align}
Thus the CFIM of MLE method is given as:
\begin{equation}
\label{eq:a1-6}
	\mathbb{F}^{(M)} = \left(
	\begin{array}{cc}
	(2n+1)(\mathrm{Re}A_{w})^{2}\sigma^{-2} & 2(\mathrm{Re}A_{w})(\mathrm{Im}A_{w})\\
	2(\mathrm{Re}A_{w})(\mathrm{Im}A_{w}) & 4(2n+1)(\mathrm{Im}A_{w})^{2}\sigma^{2}
	\end{array}
\right)
\end{equation}
If there are $N$ original samples, the efficient sample number should reduce to $N'=|\langle f|i\rangle|^{2}N$ because of pre- and post-selection. Thus $N'$ time CFIM of MLE method is $\mathbb{F}^{(M)}_{N'}=N'\mathbb{F}^{(M)}=|\langle f|i\rangle|^{2}N\mathbb{F}^{(M)}$, i.e. the result we show in Eq.(\ref{eq:3-3}).

\section{Tradeoff Relation in Homodyne Detection Strategy} \label{app:2}
Because $\mathrm{Tr}[\mathbb{F}_{N'}^{(H)}(\mathbb{Q}_{N'})^{-1}]=[\mathbb{F}_{N'}^{(H)}]_{11}/[\mathbb{Q}_{N'}]_{11}+[\mathbb{F}_{N'}^{(H)}]_{22}/[\mathbb{Q}_{N'}]_{22}$, we only care about Fisher information of $d$ and $k$ in this section. Plugging Eq.(\ref{eq:2-5}) and Eq.(\ref{eq:3-4a}) into Eq.(\ref{eq:sup3-1a}), then combining the result with Eq.(\ref{eq:3-5}), we calculate the intensity strengths of two ports in detector 1:
\begin{eqnarray}
	I_{H1}^{+} = &&\frac{1}{2}\left(A_{f1}^{2}+A_{LO1}^{2}\right)+A_{f1}A_{LO1}\frac{A_{w}}{2\sigma}\left[\sqrt{n} \alpha_{1}^{*} \left(d+2 i k \sigma ^2\right)-\sqrt{n+1}\beta_{1}^{*} \left(d-2 i k \sigma ^2\right)\right] \nonumber\\
	&&+A_{f1}A_{LO1}\frac{A_{w}^{*}d}{2\sigma}\left(\alpha_{1} \sqrt{n}-\beta_{1} \sqrt{n+1}\right)-iA_{f1}A_{LO1}k\sigma\left(\alpha_{1} \sqrt{n}+\beta_{1} \sqrt{n+1}\right) \label{eq:a2-1}
\end{eqnarray}
\begin{eqnarray}
	I_{H1}^{-} = &&\frac{1}{2}\left(A_{f1}^{2}+A_{LO1}^{2}\right)-A_{f1}A_{LO1}\frac{A_{w}}{2\sigma}\left[\sqrt{n} \alpha_{1}^{*} \left(d+2 i k \sigma ^2\right)-\sqrt{n+1}\beta_{1}^{*} \left(d-2 i k \sigma ^2\right)\right] \nonumber\\
	&&-A_{f1}A_{LO1}\frac{A_{w}^{*}d}{2\sigma}\left(\alpha_{1} \sqrt{n}-\beta_{1} \sqrt{n+1}\right)+iA_{f1}A_{LO1}k\sigma\left(\alpha_{1} \sqrt{n}+\beta_{1} \sqrt{n+1}\right) \label{eq:a2-2}
\end{eqnarray}
Similarly, the intensity strengths of two ports in detector 2 is:
\begin{eqnarray}
	I_{H2}^{+} = &&\frac{1}{2}\left(A_{f2}^{2}+A_{LO2}^{2}\right)+A_{f2}A_{LO2}\frac{A_{w}}{2\sigma}\left[\sqrt{n} \alpha_{2}^{*} \left(d+2 i k \sigma ^2\right)-\sqrt{n+1}\beta_{2}^{*} \left(d-2 i k \sigma ^2\right)\right] \nonumber\\
	&&+A_{f2}A_{LO2}\frac{A_{w}^{*}d}{2\sigma}\left(\alpha_{2} \sqrt{n}-\beta_{2} \sqrt{n+1}\right)-iA_{f2}A_{LO2}k\sigma\left(\alpha_{2} \sqrt{n}+\beta_{2} \sqrt{n+1}\right) \label{eq:a2-3}
\end{eqnarray}
\begin{eqnarray}
	I_{H2}^{-} = &&\frac{1}{2}\left(A_{f2}^{2}+A_{LO2}^{2}\right)-A_{f1}A_{LO1}\frac{A_{w}}{2\sigma}\left[\sqrt{n} \alpha_{2}^{*} \left(d+2 i k \sigma ^2\right)-\sqrt{n+1}\beta_{2}^{*} \left(d-2 i k \sigma ^2\right)\right] \nonumber\\
	&&-A_{f2}A_{LO2}\frac{A_{w}^{*}d}{2\sigma}\left(\alpha_{2} \sqrt{n}-\beta_{2} \sqrt{n+1}\right)+iA_{f2}A_{LO2}k\sigma\left(\alpha_{2} \sqrt{n}+\beta_{2} \sqrt{n+1}\right) \label{eq:a2-4}
\end{eqnarray}

Combining Eq.(\ref{eq:a2-1}) to Eq.(\ref{eq:a2-4}) with Eq.(\ref{eq:3-7}) and Eq.(\ref{eq:3-8}), and we let $A_{f}=\sqrt{2}A_{f1}=\sqrt{2}A_{f2}=\sqrt{N'}$, $A_{LO}=\sqrt{2}A_{LO1}=\sqrt{2}A_{LO2}=\sqrt{N_{LO}}$, where $N'=|\langle f|i\rangle|^{2}N$. Considering $N_{LO}\gg N'$, we can get Fisher information about parameter $d$:
\begin{equation}
	\label{eq:a2-5}
	[\mathbb{F}_{N'}^{(H)}]_{11}=\frac{|\langle f|i\rangle|^{2}N}{4\sigma^{2}}\sum_{i=1,2}n|A_{w}^{*}\alpha_{i}|^{2}+(n+1)|A_{w}^{*}\beta_{i}|^{2}-2\sqrt{n(n+1)}\mathrm{Re}[(A_{w}^{*})^{2}\alpha_{i}\beta_{i}]
\end{equation}
and the Fisher information about parameter $k$ can be also derived:
\begin{equation}
\label{eq:a2-6}
	[\mathbb{F}_{N'}^{(H)}]_{22}=|\langle f|i\rangle|^{2}N\sigma^{2}\sum_{i=1,2}n|A_{w}^{*}\alpha_{i}|^{2}+(n+1)|A_{w}^{*}\beta_{i}|^{2}-2\sqrt{n(n+1)}\mathrm{Re}[(A_{w}^{*})^{2}\alpha_{i}\beta_{i}]
\end{equation}

Thus the tradeoff relation in homodyne detection strategy is:
\begin{eqnarray}
	\mathrm{Tr}[\mathbb{F}_{N'}^{(H)}(\mathbb{Q}_{N'})^{-1}]=&&[\mathbb{F}_{N'}^{(H)}]_{11}/[\mathbb{Q}_{N'}]_{11}+[\mathbb{F}_{N'}^{(H)}]_{22}/[\mathbb{Q}_{N'}]_{22} \nonumber\\
	=&&\frac{1}{2(2n+1)|A_{w}|^{2}}\sum_{i=1,2}n|A_{w}^{*}\alpha_{i}|^{2}+(n+1)|A_{w}^{*}\beta_{i}|^{2}-2\sqrt{n(n+1)}\mathrm{Re}[(A_{w}^{*})^{2}\alpha_{i}\beta_{i}] \nonumber\\
	=&&\sum_{i=1,2}\frac{n|A_{w}^{*}\alpha_{i}|^{2}+(n+1)|A_{w}^{*}\beta_{i}|^{2}-2\sqrt{n(n+1)}\mathrm{Re}[(A_{w}^{*})^{2}\alpha_{i}\beta_{i}]}{2(2n+1)|A_{w}^{*}|^{2}(|\alpha_{i}|^{2}+|\beta_{i}|^{2})}\quad(\text{where }|\alpha_{i}|^{2}+|\beta_{i}|^{2}=1) \label{a2-7}
\end{eqnarray}
Obviously, $\mathrm{Re}[(A_{w}^{*})^{2}\alpha_{i}\beta_{i}]\le |(A_{w}^{*})^{2}\alpha_{i}\beta_{i}|$. Using vector inequality $|\vec{\mathbf{a}}\cdot\vec{\mathbf{b}}|\le|\vec{\mathbf{a}}|\cdot|\vec{\mathbf{b}}|$, we can get
\begin{align}
	n|A_{w}^{*}\alpha_{i}|^{2}+(n+1)|A_{w}^{*}\beta_{i}|^{2}-2\sqrt{n(n+1)}|(A_{w}^{*})^{2}\alpha_{i}\beta_{i}| =& (\sqrt{n}|A_{w}^{*}\alpha|+\sqrt{n+1}|A_{w}^{*}\beta|)^{2} \nonumber\\
	\le& (n+n+1)(|A_{w}^{*}\alpha|^{2}+|A_{w}^{*}\beta|^{2}) \nonumber\\
	=&(2n+1)|A_{w}^{*}|^{2}(|\alpha_{i}|^{2}+|\beta_{i}|^{2}) \nonumber
\end{align}

Thus inequality $\mathrm{Tr}[\mathbb{F}_{N'}^{(H)}(\mathbb{Q}_{N'})^{-1}]\le 1$ is always true, and the equality relation hold if and only if
\begin{equation}
	\label{eq:a2-8}
	\left\{
	\begin{array}{ll}
	\mathrm{Re}(A_{w}^{2}\alpha_{i}^{*}\beta_{i}^{*})=-|A_{w}\alpha_{i}^{*}|\cdot|A_{w}\beta_{i}^{*}|\\
	|\alpha_{i}|/|\beta_{i}|=\sqrt{n}/\sqrt{n+1}
	\end{array}\right.
	\qquad i=1,2
\end{equation}

\end{widetext}
\bibliography{main}

\begin{thebibliography}{35}%
\makeatletter
\providecommand \@ifxundefined [1]{%
 \@ifx{#1\undefined}
}%
\providecommand \@ifnum [1]{%
 \ifnum #1\expandafter \@firstoftwo
 \else \expandafter \@secondoftwo
 \fi
}%
\providecommand \@ifx [1]{%
 \ifx #1\expandafter \@firstoftwo
 \else \expandafter \@secondoftwo
 \fi
}%
\providecommand \natexlab [1]{#1}%
\providecommand \enquote  [1]{``#1''}%
\providecommand \bibnamefont  [1]{#1}%
\providecommand \bibfnamefont [1]{#1}%
\providecommand \citenamefont [1]{#1}%
\providecommand \href@noop [0]{\@secondoftwo}%
\providecommand \href [0]{\begingroup \@sanitize@url \@href}%
\providecommand \@href[1]{\@@startlink{#1}\@@href}%
\providecommand \@@href[1]{\endgroup#1\@@endlink}%
\providecommand \@sanitize@url [0]{\catcode `\\12\catcode `\$12\catcode
  `\&12\catcode `\#12\catcode `\^12\catcode `\_12\catcode `\%12\relax}%
\providecommand \@@startlink[1]{}%
\providecommand \@@endlink[0]{}%
\providecommand \url  [0]{\begingroup\@sanitize@url \@url }%
\providecommand \@url [1]{\endgroup\@href {#1}{\urlprefix }}%
\providecommand \urlprefix  [0]{URL }%
\providecommand \Eprint [0]{\href }%
\providecommand \doibase [0]{https://doi.org/}%
\providecommand \selectlanguage [0]{\@gobble}%
\providecommand \bibinfo  [0]{\@secondoftwo}%
\providecommand \bibfield  [0]{\@secondoftwo}%
\providecommand \translation [1]{[#1]}%
\providecommand \BibitemOpen [0]{}%
\providecommand \bibitemStop [0]{}%
\providecommand \bibitemNoStop [0]{.\EOS\space}%
\providecommand \EOS [0]{\spacefactor3000\relax}%
\providecommand \BibitemShut  [1]{\csname bibitem#1\endcsname}%
\let\auto@bib@innerbib\@empty
\bibitem [{\citenamefont {Aharonov}\ \emph {et~al.}(1988)\citenamefont
  {Aharonov}, \citenamefont {Albert},\ and\ \citenamefont
  {Vaidman}}]{PhysRevLett.60.1351}%
  \BibitemOpen
  \bibfield  {author} {\bibinfo {author} {\bibfnamefont {Y.}~\bibnamefont
  {Aharonov}}, \bibinfo {author} {\bibfnamefont {D.~Z.}\ \bibnamefont
  {Albert}},\ and\ \bibinfo {author} {\bibfnamefont {L.}~\bibnamefont
  {Vaidman}},\ }\bibfield  {title} {\bibinfo {title} {How the result of a
  measurement of a component of the spin of a spin-1/2 particle can turn out to
  be 100},\ }\href {https://doi.org/10.1103/PhysRevLett.60.1351} {\bibfield
  {journal} {\bibinfo  {journal} {Phys. Rev. Lett.}\ }\textbf {\bibinfo
  {volume} {60}},\ \bibinfo {pages} {1351} (\bibinfo {year}
  {1988})}\BibitemShut {NoStop}%
\bibitem [{\citenamefont {Ritchie}\ \emph {et~al.}(1991)\citenamefont
  {Ritchie}, \citenamefont {Story},\ and\ \citenamefont
  {Hulet}}]{PhysRevLett.66.1107}%
  \BibitemOpen
  \bibfield  {author} {\bibinfo {author} {\bibfnamefont {N.~W.~M.}\
  \bibnamefont {Ritchie}}, \bibinfo {author} {\bibfnamefont {J.~G.}\
  \bibnamefont {Story}},\ and\ \bibinfo {author} {\bibfnamefont {R.~G.}\
  \bibnamefont {Hulet}},\ }\bibfield  {title} {\bibinfo {title} {Realization of
  a measurement of a ``weak value''},\ }\href
  {https://doi.org/10.1103/PhysRevLett.66.1107} {\bibfield  {journal} {\bibinfo
   {journal} {Phys. Rev. Lett.}\ }\textbf {\bibinfo {volume} {66}},\ \bibinfo
  {pages} {1107} (\bibinfo {year} {1991})}\BibitemShut {NoStop}%
\bibitem [{\citenamefont {Hosten}\ and\ \citenamefont
  {Kwiat}(2008)}]{Hosten787}%
  \BibitemOpen
  \bibfield  {author} {\bibinfo {author} {\bibfnamefont {O.}~\bibnamefont
  {Hosten}}\ and\ \bibinfo {author} {\bibfnamefont {P.}~\bibnamefont {Kwiat}},\
  }\bibfield  {title} {\bibinfo {title} {Observation of the spin hall effect of
  light via weak measurements},\ }\href
  {https://doi.org/10.1126/science.1152697} {\bibfield  {journal} {\bibinfo
  {journal} {Science}\ }\textbf {\bibinfo {volume} {319}},\ \bibinfo {pages}
  {787} (\bibinfo {year} {2008})}\BibitemShut {NoStop}%
\bibitem [{\citenamefont {Dixon}\ \emph {et~al.}(2009)\citenamefont {Dixon},
  \citenamefont {Starling}, \citenamefont {Jordan},\ and\ \citenamefont
  {Howell}}]{PhysRevLett.102.173601}%
  \BibitemOpen
  \bibfield  {author} {\bibinfo {author} {\bibfnamefont {P.~B.}\ \bibnamefont
  {Dixon}}, \bibinfo {author} {\bibfnamefont {D.~J.}\ \bibnamefont {Starling}},
  \bibinfo {author} {\bibfnamefont {A.~N.}\ \bibnamefont {Jordan}},\ and\
  \bibinfo {author} {\bibfnamefont {J.~C.}\ \bibnamefont {Howell}},\ }\bibfield
   {title} {\bibinfo {title} {Ultrasensitive beam deflection measurement via
  interferometric weak value amplification},\ }\href
  {https://doi.org/10.1103/PhysRevLett.102.173601} {\bibfield  {journal}
  {\bibinfo  {journal} {Phys. Rev. Lett.}\ }\textbf {\bibinfo {volume} {102}},\
  \bibinfo {pages} {173601} (\bibinfo {year} {2009})}\BibitemShut {NoStop}%
\bibitem [{\citenamefont {Zhang}\ \emph {et~al.}(2015)\citenamefont {Zhang},
  \citenamefont {Datta},\ and\ \citenamefont
  {Walmsley}}]{PhysRevLett.114.210801}%
  \BibitemOpen
  \bibfield  {author} {\bibinfo {author} {\bibfnamefont {L.}~\bibnamefont
  {Zhang}}, \bibinfo {author} {\bibfnamefont {A.}~\bibnamefont {Datta}},\ and\
  \bibinfo {author} {\bibfnamefont {I.~A.}\ \bibnamefont {Walmsley}},\
  }\bibfield  {title} {\bibinfo {title} {Precision metrology using weak
  measurements},\ }\href {https://doi.org/10.1103/PhysRevLett.114.210801}
  {\bibfield  {journal} {\bibinfo  {journal} {Phys. Rev. Lett.}\ }\textbf
  {\bibinfo {volume} {114}},\ \bibinfo {pages} {210801} (\bibinfo {year}
  {2015})}\BibitemShut {NoStop}%
\bibitem [{\citenamefont {Fang}\ \emph {et~al.}(2018)\citenamefont {Fang},
  \citenamefont {Huang},\ and\ \citenamefont {Zeng}}]{PhysRevA.97.063818}%
  \BibitemOpen
  \bibfield  {author} {\bibinfo {author} {\bibfnamefont {C.}~\bibnamefont
  {Fang}}, \bibinfo {author} {\bibfnamefont {J.-Z.}\ \bibnamefont {Huang}},\
  and\ \bibinfo {author} {\bibfnamefont {G.}~\bibnamefont {Zeng}},\ }\bibfield
  {title} {\bibinfo {title} {Robust interferometry against imperfections based
  on weak value amplification},\ }\href
  {https://doi.org/10.1103/PhysRevA.97.063818} {\bibfield  {journal} {\bibinfo
  {journal} {Phys. Rev. A}\ }\textbf {\bibinfo {volume} {97}},\ \bibinfo
  {pages} {063818} (\bibinfo {year} {2018})}\BibitemShut {NoStop}%
\bibitem [{\citenamefont {Fang}\ \emph {et~al.}(2016)\citenamefont {Fang},
  \citenamefont {Huang}, \citenamefont {Yu}, \citenamefont {Li},\ and\
  \citenamefont {Zeng}}]{Fang_2016}%
  \BibitemOpen
  \bibfield  {author} {\bibinfo {author} {\bibfnamefont {C.}~\bibnamefont
  {Fang}}, \bibinfo {author} {\bibfnamefont {J.-Z.}\ \bibnamefont {Huang}},
  \bibinfo {author} {\bibfnamefont {Y.}~\bibnamefont {Yu}}, \bibinfo {author}
  {\bibfnamefont {Q.}~\bibnamefont {Li}},\ and\ \bibinfo {author}
  {\bibfnamefont {G.}~\bibnamefont {Zeng}},\ }\bibfield  {title} {\bibinfo
  {title} {Ultra-small time-delay estimation via a weak measurement technique
  with post-selection},\ }\href
  {https://doi.org/10.1088/0953-4075/49/17/175501} {\bibfield  {journal}
  {\bibinfo  {journal} {Journal of Physics B: Atomic, Molecular and Optical
  Physics}\ }\textbf {\bibinfo {volume} {49}},\ \bibinfo {pages} {175501}
  (\bibinfo {year} {2016})}\BibitemShut {NoStop}%
\bibitem [{\citenamefont {Huang}\ \emph {et~al.}(2018)\citenamefont {Huang},
  \citenamefont {Fang},\ and\ \citenamefont {Zeng}}]{PhysRevA.97.063853}%
  \BibitemOpen
  \bibfield  {author} {\bibinfo {author} {\bibfnamefont {J.-Z.}\ \bibnamefont
  {Huang}}, \bibinfo {author} {\bibfnamefont {C.}~\bibnamefont {Fang}},\ and\
  \bibinfo {author} {\bibfnamefont {G.}~\bibnamefont {Zeng}},\ }\bibfield
  {title} {\bibinfo {title} {Weak-value-amplification metrology without
  spectral analysis},\ }\href {https://doi.org/10.1103/PhysRevA.97.063853}
  {\bibfield  {journal} {\bibinfo  {journal} {Phys. Rev. A}\ }\textbf {\bibinfo
  {volume} {97}},\ \bibinfo {pages} {063853} (\bibinfo {year}
  {2018})}\BibitemShut {NoStop}%
\bibitem [{\citenamefont {Feizpour}\ \emph {et~al.}(2011)\citenamefont
  {Feizpour}, \citenamefont {Xing},\ and\ \citenamefont
  {Steinberg}}]{PhysRevLett.107.133603}%
  \BibitemOpen
  \bibfield  {author} {\bibinfo {author} {\bibfnamefont {A.}~\bibnamefont
  {Feizpour}}, \bibinfo {author} {\bibfnamefont {X.}~\bibnamefont {Xing}},\
  and\ \bibinfo {author} {\bibfnamefont {A.~M.}\ \bibnamefont {Steinberg}},\
  }\bibfield  {title} {\bibinfo {title} {Amplifying single-photon nonlinearity
  using weak measurements},\ }\href
  {https://doi.org/10.1103/PhysRevLett.107.133603} {\bibfield  {journal}
  {\bibinfo  {journal} {Phys. Rev. Lett.}\ }\textbf {\bibinfo {volume} {107}},\
  \bibinfo {pages} {133603} (\bibinfo {year} {2011})}\BibitemShut {NoStop}%
\bibitem [{\citenamefont {Hallaji}\ \emph {et~al.}(2017)\citenamefont
  {Hallaji}, \citenamefont {Feizpour}, \citenamefont {Dmochowski},
  \citenamefont {Sinclair},\ and\ \citenamefont {Steinberg}}]{nphys4040}%
  \BibitemOpen
  \bibfield  {author} {\bibinfo {author} {\bibfnamefont {M.}~\bibnamefont
  {Hallaji}}, \bibinfo {author} {\bibfnamefont {A.}~\bibnamefont {Feizpour}},
  \bibinfo {author} {\bibfnamefont {G.}~\bibnamefont {Dmochowski}}, \bibinfo
  {author} {\bibfnamefont {J.}~\bibnamefont {Sinclair}},\ and\ \bibinfo
  {author} {\bibfnamefont {A.}~\bibnamefont {Steinberg}},\ }\bibfield  {title}
  {\bibinfo {title} {Weak-value amplification of the nonlinear effect of a
  single photon},\ }\href {https://doi.org/10.1038/nphys4040} {\bibfield
  {journal} {\bibinfo  {journal} {Nature Physics}\ }\textbf {\bibinfo {volume}
  {13}},\ \bibinfo {pages} {540} (\bibinfo {year} {2017})}\BibitemShut
  {NoStop}%
\bibitem [{\citenamefont {Li}\ \emph {et~al.}(2018)\citenamefont {Li},
  \citenamefont {Huang}, \citenamefont {Yu}, \citenamefont {Li}, \citenamefont
  {Fang},\ and\ \citenamefont {Zeng}}]{doi:10.1063/1.5027117}%
  \BibitemOpen
  \bibfield  {author} {\bibinfo {author} {\bibfnamefont {H.}~\bibnamefont
  {Li}}, \bibinfo {author} {\bibfnamefont {J.-Z.}\ \bibnamefont {Huang}},
  \bibinfo {author} {\bibfnamefont {Y.}~\bibnamefont {Yu}}, \bibinfo {author}
  {\bibfnamefont {Y.}~\bibnamefont {Li}}, \bibinfo {author} {\bibfnamefont
  {C.}~\bibnamefont {Fang}},\ and\ \bibinfo {author} {\bibfnamefont
  {G.}~\bibnamefont {Zeng}},\ }\bibfield  {title} {\bibinfo {title}
  {High-precision temperature measurement based on weak measurement using
  nematic liquid crystals},\ }\href {https://doi.org/10.1063/1.5027117}
  {\bibfield  {journal} {\bibinfo  {journal} {Applied Physics Letters}\
  }\textbf {\bibinfo {volume} {112}},\ \bibinfo {pages} {231901} (\bibinfo
  {year} {2018})}\BibitemShut {NoStop}%
\bibitem [{\citenamefont {Brunner}\ and\ \citenamefont
  {Simon}(2010)}]{PhysRevLett.105.010405}%
  \BibitemOpen
  \bibfield  {author} {\bibinfo {author} {\bibfnamefont {N.}~\bibnamefont
  {Brunner}}\ and\ \bibinfo {author} {\bibfnamefont {C.}~\bibnamefont
  {Simon}},\ }\bibfield  {title} {\bibinfo {title} {Measuring small
  longitudinal phase shifts: Weak measurements or standard interferometry?},\
  }\href {https://doi.org/10.1103/PhysRevLett.105.010405} {\bibfield  {journal}
  {\bibinfo  {journal} {Phys. Rev. Lett.}\ }\textbf {\bibinfo {volume} {105}},\
  \bibinfo {pages} {010405} (\bibinfo {year} {2010})}\BibitemShut {NoStop}%
\bibitem [{\citenamefont {Xu}\ \emph {et~al.}(2013)\citenamefont {Xu},
  \citenamefont {Kedem}, \citenamefont {Sun}, \citenamefont {Vaidman},
  \citenamefont {Li},\ and\ \citenamefont {Guo}}]{PhysRevLett.111.033604}%
  \BibitemOpen
  \bibfield  {author} {\bibinfo {author} {\bibfnamefont {X.-Y.}\ \bibnamefont
  {Xu}}, \bibinfo {author} {\bibfnamefont {Y.}~\bibnamefont {Kedem}}, \bibinfo
  {author} {\bibfnamefont {K.}~\bibnamefont {Sun}}, \bibinfo {author}
  {\bibfnamefont {L.}~\bibnamefont {Vaidman}}, \bibinfo {author} {\bibfnamefont
  {C.-F.}\ \bibnamefont {Li}},\ and\ \bibinfo {author} {\bibfnamefont {G.-C.}\
  \bibnamefont {Guo}},\ }\bibfield  {title} {\bibinfo {title} {Phase estimation
  with weak measurement using a white light source},\ }\href
  {https://doi.org/10.1103/PhysRevLett.111.033604} {\bibfield  {journal}
  {\bibinfo  {journal} {Phys. Rev. Lett.}\ }\textbf {\bibinfo {volume} {111}},\
  \bibinfo {pages} {033604} (\bibinfo {year} {2013})}\BibitemShut {NoStop}%
\bibitem [{\citenamefont {Dressel}\ \emph {et~al.}(2014)\citenamefont
  {Dressel}, \citenamefont {Malik}, \citenamefont {Miatto}, \citenamefont
  {Jordan},\ and\ \citenamefont {Boyd}}]{RevModPhys.86.307}%
  \BibitemOpen
  \bibfield  {author} {\bibinfo {author} {\bibfnamefont {J.}~\bibnamefont
  {Dressel}}, \bibinfo {author} {\bibfnamefont {M.}~\bibnamefont {Malik}},
  \bibinfo {author} {\bibfnamefont {F.~M.}\ \bibnamefont {Miatto}}, \bibinfo
  {author} {\bibfnamefont {A.~N.}\ \bibnamefont {Jordan}},\ and\ \bibinfo
  {author} {\bibfnamefont {R.~W.}\ \bibnamefont {Boyd}},\ }\bibfield  {title}
  {\bibinfo {title} {Colloquium: Understanding quantum weak values: Basics and
  applications},\ }\href {https://doi.org/10.1103/RevModPhys.86.307} {\bibfield
   {journal} {\bibinfo  {journal} {Rev. Mod. Phys.}\ }\textbf {\bibinfo
  {volume} {86}},\ \bibinfo {pages} {307} (\bibinfo {year} {2014})}\BibitemShut
  {NoStop}%
\bibitem [{\citenamefont {Dziewior}\ \emph {et~al.}(2019)\citenamefont
  {Dziewior}, \citenamefont {Knips}, \citenamefont {Farfurnik}, \citenamefont
  {Senkalla}, \citenamefont {Benshalom}, \citenamefont {Efroni}, \citenamefont
  {Meinecke}, \citenamefont {Bar-Ad}, \citenamefont {Weinfurter},\ and\
  \citenamefont {Vaidman}}]{Dziewior2881}%
  \BibitemOpen
  \bibfield  {author} {\bibinfo {author} {\bibfnamefont {J.}~\bibnamefont
  {Dziewior}}, \bibinfo {author} {\bibfnamefont {L.}~\bibnamefont {Knips}},
  \bibinfo {author} {\bibfnamefont {D.}~\bibnamefont {Farfurnik}}, \bibinfo
  {author} {\bibfnamefont {K.}~\bibnamefont {Senkalla}}, \bibinfo {author}
  {\bibfnamefont {N.}~\bibnamefont {Benshalom}}, \bibinfo {author}
  {\bibfnamefont {J.}~\bibnamefont {Efroni}}, \bibinfo {author} {\bibfnamefont
  {J.}~\bibnamefont {Meinecke}}, \bibinfo {author} {\bibfnamefont
  {S.}~\bibnamefont {Bar-Ad}}, \bibinfo {author} {\bibfnamefont
  {H.}~\bibnamefont {Weinfurter}},\ and\ \bibinfo {author} {\bibfnamefont
  {L.}~\bibnamefont {Vaidman}},\ }\bibfield  {title} {\bibinfo {title}
  {Universality of local weak interactions and its application for
  interferometric alignment},\ }\href {https://doi.org/10.1073/pnas.1812970116}
  {\bibfield  {journal} {\bibinfo  {journal} {Proceedings of the National
  Academy of Sciences}\ }\textbf {\bibinfo {volume} {116}},\ \bibinfo {pages}
  {2881} (\bibinfo {year} {2019})}\BibitemShut {NoStop}%
\bibitem [{\citenamefont {Vella}\ \emph {et~al.}(2019)\citenamefont {Vella},
  \citenamefont {Head}, \citenamefont {Brown},\ and\ \citenamefont
  {Alonso}}]{PhysRevLett.122.123603}%
  \BibitemOpen
  \bibfield  {author} {\bibinfo {author} {\bibfnamefont {A.}~\bibnamefont
  {Vella}}, \bibinfo {author} {\bibfnamefont {S.~T.}\ \bibnamefont {Head}},
  \bibinfo {author} {\bibfnamefont {T.~G.}\ \bibnamefont {Brown}},\ and\
  \bibinfo {author} {\bibfnamefont {M.~A.}\ \bibnamefont {Alonso}},\ }\bibfield
   {title} {\bibinfo {title} {Simultaneous measurement of multiple parameters
  of a subwavelength structure based on the weak value formalism},\ }\href
  {https://doi.org/10.1103/PhysRevLett.122.123603} {\bibfield  {journal}
  {\bibinfo  {journal} {Phys. Rev. Lett.}\ }\textbf {\bibinfo {volume} {122}},\
  \bibinfo {pages} {123603} (\bibinfo {year} {2019})}\BibitemShut {NoStop}%
\bibitem [{\citenamefont {{Ho}}\ and\ \citenamefont
  {{Kondo}}(2018)}]{2018arXiv181108046H}%
  \BibitemOpen
  \bibfield  {author} {\bibinfo {author} {\bibfnamefont {L.~B.}\ \bibnamefont
  {{Ho}}}\ and\ \bibinfo {author} {\bibfnamefont {Y.}~\bibnamefont {{Kondo}}},\
  }\bibfield  {title} {\bibinfo {title} {{Tradeoffs in multiple-parameter
  postselection measurements}},\ }\href@noop {} {\bibfield  {journal} {\bibinfo
   {journal} {arXiv e-prints}\ ,\ \bibinfo {eid} {arXiv:1811.08046}} (\bibinfo
  {year} {2018})},\ \Eprint {https://arxiv.org/abs/1811.08046}
  {arXiv:1811.08046 [quant-ph]} \BibitemShut {NoStop}%
\bibitem [{\citenamefont {Jordan}\ \emph {et~al.}(2014)\citenamefont {Jordan},
  \citenamefont {Mart\'{\i}nez-Rinc\'on},\ and\ \citenamefont
  {Howell}}]{PhysRevX.4.011031}%
  \BibitemOpen
  \bibfield  {author} {\bibinfo {author} {\bibfnamefont {A.~N.}\ \bibnamefont
  {Jordan}}, \bibinfo {author} {\bibfnamefont {J.}~\bibnamefont
  {Mart\'{\i}nez-Rinc\'on}},\ and\ \bibinfo {author} {\bibfnamefont {J.~C.}\
  \bibnamefont {Howell}},\ }\bibfield  {title} {\bibinfo {title} {Technical
  advantages for weak-value amplification: When less is more},\ }\href
  {https://doi.org/10.1103/PhysRevX.4.011031} {\bibfield  {journal} {\bibinfo
  {journal} {Phys. Rev. X}\ }\textbf {\bibinfo {volume} {4}},\ \bibinfo {pages}
  {011031} (\bibinfo {year} {2014})}\BibitemShut {NoStop}%
\bibitem [{\citenamefont {{Mori}}\ \emph {et~al.}(2019)\citenamefont {{Mori}},
  \citenamefont {{Lee}},\ and\ \citenamefont
  {{Tsutsui}}}]{2019arXiv190106831M}%
  \BibitemOpen
  \bibfield  {author} {\bibinfo {author} {\bibfnamefont {Y.}~\bibnamefont
  {{Mori}}}, \bibinfo {author} {\bibfnamefont {J.}~\bibnamefont {{Lee}}},\ and\
  \bibinfo {author} {\bibfnamefont {I.}~\bibnamefont {{Tsutsui}}},\ }\bibfield
  {title} {\bibinfo {title} {{On the Validity of Weak Measurement Applied for
  Precision Measurement}},\ }\href@noop {} {\bibfield  {journal} {\bibinfo
  {journal} {arXiv e-prints}\ ,\ \bibinfo {eid} {arXiv:1901.06831}} (\bibinfo
  {year} {2019})},\ \Eprint {https://arxiv.org/abs/1901.06831}
  {arXiv:1901.06831} \BibitemShut {NoStop}%
\bibitem [{\citenamefont {Aharonov}\ and\ \citenamefont
  {Vaidman}(1990)}]{PhysRevA.41.11}%
  \BibitemOpen
  \bibfield  {author} {\bibinfo {author} {\bibfnamefont {Y.}~\bibnamefont
  {Aharonov}}\ and\ \bibinfo {author} {\bibfnamefont {L.}~\bibnamefont
  {Vaidman}},\ }\bibfield  {title} {\bibinfo {title} {Properties of a quantum
  system during the time interval between two measurements},\ }\href
  {https://doi.org/10.1103/PhysRevA.41.11} {\bibfield  {journal} {\bibinfo
  {journal} {Phys. Rev. A}\ }\textbf {\bibinfo {volume} {41}},\ \bibinfo
  {pages} {11} (\bibinfo {year} {1990})}\BibitemShut {NoStop}%
\bibitem [{\citenamefont {Howell}\ \emph {et~al.}(2010)\citenamefont {Howell},
  \citenamefont {Starling}, \citenamefont {Dixon}, \citenamefont {Vudyasetu},\
  and\ \citenamefont {Jordan}}]{PhysRevA.81.033813}%
  \BibitemOpen
  \bibfield  {author} {\bibinfo {author} {\bibfnamefont {J.~C.}\ \bibnamefont
  {Howell}}, \bibinfo {author} {\bibfnamefont {D.~J.}\ \bibnamefont
  {Starling}}, \bibinfo {author} {\bibfnamefont {P.~B.}\ \bibnamefont {Dixon}},
  \bibinfo {author} {\bibfnamefont {P.~K.}\ \bibnamefont {Vudyasetu}},\ and\
  \bibinfo {author} {\bibfnamefont {A.~N.}\ \bibnamefont {Jordan}},\ }\bibfield
   {title} {\bibinfo {title} {Interferometric weak value deflections: Quantum
  and classical treatments},\ }\href
  {https://doi.org/10.1103/PhysRevA.81.033813} {\bibfield  {journal} {\bibinfo
  {journal} {Phys. Rev. A}\ }\textbf {\bibinfo {volume} {81}},\ \bibinfo
  {pages} {033813} (\bibinfo {year} {2010})}\BibitemShut {NoStop}%
\bibitem [{\citenamefont {Helstrom}(1969)}]{Helstrom1969}%
  \BibitemOpen
  \bibfield  {author} {\bibinfo {author} {\bibfnamefont {C.~W.}\ \bibnamefont
  {Helstrom}},\ }\bibfield  {title} {\bibinfo {title} {Quantum detection and
  estimation theory},\ }\href {https://doi.org/10.1007/BF01007479} {\bibfield
  {journal} {\bibinfo  {journal} {Journal of Statistical Physics}\ }\textbf
  {\bibinfo {volume} {1}},\ \bibinfo {pages} {231} (\bibinfo {year}
  {1969})}\BibitemShut {NoStop}%
\bibitem [{\citenamefont {Holevo}(2011)}]{holevo2011probabilistic}%
  \BibitemOpen
  \bibfield  {author} {\bibinfo {author} {\bibfnamefont {A.}~\bibnamefont
  {Holevo}},\ }\href {https://books.google.co.jp/books?id=l7AIDhbWrTIC} {\emph
  {\bibinfo {title} {Probabilistic and Statistical Aspects of Quantum
  Theory}}},\ Publications of the Scuola Normale Superiore\ (\bibinfo
  {publisher} {Scuola Normale Superiore},\ \bibinfo {year} {2011})\BibitemShut
  {NoStop}%
\bibitem [{\citenamefont {Demkowicz-Dobrzański}\ \emph
  {et~al.}(2015)\citenamefont {Demkowicz-Dobrzański}, \citenamefont
  {Jarzyna},\ and\ \citenamefont {Kołodyński}}]{DEMKOWICZDOBRZANSKI2015345}%
  \BibitemOpen
  \bibfield  {author} {\bibinfo {author} {\bibfnamefont {R.}~\bibnamefont
  {Demkowicz-Dobrzański}}, \bibinfo {author} {\bibfnamefont {M.}~\bibnamefont
  {Jarzyna}},\ and\ \bibinfo {author} {\bibfnamefont {J.}~\bibnamefont
  {Kołodyński}},\ }\bibfield  {title} {\bibinfo {title} {Chapter four -
  quantum limits in optical interferometry}\ }(\bibinfo  {publisher}
  {Elsevier},\ \bibinfo {year} {2015})\ pp.\ \bibinfo {pages} {345 --
  435}\BibitemShut {NoStop}%
\bibitem [{\citenamefont {Turek}\ \emph {et~al.}(2015)\citenamefont {Turek},
  \citenamefont {Kobayashi}, \citenamefont {Akutsu}, \citenamefont {Sun},\ and\
  \citenamefont {Shikano}}]{Turek_2015}%
  \BibitemOpen
  \bibfield  {author} {\bibinfo {author} {\bibfnamefont {Y.}~\bibnamefont
  {Turek}}, \bibinfo {author} {\bibfnamefont {H.}~\bibnamefont {Kobayashi}},
  \bibinfo {author} {\bibfnamefont {T.}~\bibnamefont {Akutsu}}, \bibinfo
  {author} {\bibfnamefont {C.-P.}\ \bibnamefont {Sun}},\ and\ \bibinfo {author}
  {\bibfnamefont {Y.}~\bibnamefont {Shikano}},\ }\bibfield  {title} {\bibinfo
  {title} {Post-selected von neumann measurement with
  hermite{\textendash}gaussian and laguerre{\textendash}gaussian pointer
  states},\ }\href {https://doi.org/10.1088/1367-2630/17/8/083029} {\bibfield
  {journal} {\bibinfo  {journal} {New Journal of Physics}\ }\textbf {\bibinfo
  {volume} {17}},\ \bibinfo {pages} {083029} (\bibinfo {year}
  {2015})}\BibitemShut {NoStop}%
\bibitem [{\citenamefont {Dressel}\ and\ \citenamefont
  {Jordan}(2012)}]{PhysRevLett.109.230402}%
  \BibitemOpen
  \bibfield  {author} {\bibinfo {author} {\bibfnamefont {J.}~\bibnamefont
  {Dressel}}\ and\ \bibinfo {author} {\bibfnamefont {A.~N.}\ \bibnamefont
  {Jordan}},\ }\bibfield  {title} {\bibinfo {title} {Weak values are universal
  in von neumann measurements},\ }\href
  {https://doi.org/10.1103/PhysRevLett.109.230402} {\bibfield  {journal}
  {\bibinfo  {journal} {Phys. Rev. Lett.}\ }\textbf {\bibinfo {volume} {109}},\
  \bibinfo {pages} {230402} (\bibinfo {year} {2012})}\BibitemShut {NoStop}%
\bibitem [{\citenamefont {{Vella}}(2018)}]{2018arXiv180604503V}%
  \BibitemOpen
  \bibfield  {author} {\bibinfo {author} {\bibfnamefont {A.}~\bibnamefont
  {{Vella}}},\ }\bibfield  {title} {\bibinfo {title} {{Tutorial: Maximum
  likelihood estimation in the context of an optical measurement}},\
  }\href@noop {} {\bibfield  {journal} {\bibinfo  {journal} {arXiv e-prints}\
  ,\ \bibinfo {eid} {arXiv:1806.04503}} (\bibinfo {year} {2018})},\ \Eprint
  {https://arxiv.org/abs/1806.04503} {arXiv:1806.04503 [physics.data-an]}
  \BibitemShut {NoStop}%
\bibitem [{\citenamefont {Li}\ \emph {et~al.}(2017)\citenamefont {Li},
  \citenamefont {Huang},\ and\ \citenamefont {Zeng}}]{PhysRevA.96.032112}%
  \BibitemOpen
  \bibfield  {author} {\bibinfo {author} {\bibfnamefont {F.}~\bibnamefont
  {Li}}, \bibinfo {author} {\bibfnamefont {J.}~\bibnamefont {Huang}},\ and\
  \bibinfo {author} {\bibfnamefont {G.}~\bibnamefont {Zeng}},\ }\bibfield
  {title} {\bibinfo {title} {Adaptive weak-value amplification with adjustable
  postselection},\ }\href {https://doi.org/10.1103/PhysRevA.96.032112}
  {\bibfield  {journal} {\bibinfo  {journal} {Phys. Rev. A}\ }\textbf {\bibinfo
  {volume} {96}},\ \bibinfo {pages} {032112} (\bibinfo {year}
  {2017})}\BibitemShut {NoStop}%
\bibitem [{\citenamefont {Liu}\ \emph {et~al.}(2017)\citenamefont {Liu},
  \citenamefont {Mart\'{i}nez-Rinc\'{o}n}, \citenamefont {Viza},\ and\
  \citenamefont {Howell}}]{Liu:17}%
  \BibitemOpen
  \bibfield  {author} {\bibinfo {author} {\bibfnamefont {W.-T.}\ \bibnamefont
  {Liu}}, \bibinfo {author} {\bibfnamefont {J.}~\bibnamefont
  {Mart\'{i}nez-Rinc\'{o}n}}, \bibinfo {author} {\bibfnamefont {G.~I.}\
  \bibnamefont {Viza}},\ and\ \bibinfo {author} {\bibfnamefont {J.~C.}\
  \bibnamefont {Howell}},\ }\bibfield  {title} {\bibinfo {title} {Anomalous
  amplification of a homodyne signal via almost-balanced weak values},\ }\href
  {https://doi.org/10.1364/OL.42.000903} {\bibfield  {journal} {\bibinfo
  {journal} {Opt. Lett.}\ }\textbf {\bibinfo {volume} {42}},\ \bibinfo {pages}
  {903} (\bibinfo {year} {2017})}\BibitemShut {NoStop}%
\bibitem [{\citenamefont {Delaubert}\ \emph {et~al.}(2006)\citenamefont
  {Delaubert}, \citenamefont {Treps}, \citenamefont {Lassen}, \citenamefont
  {Harb}, \citenamefont {Fabre}, \citenamefont {Lam},\ and\ \citenamefont
  {Bachor}}]{PhysRevA.74.053823}%
  \BibitemOpen
  \bibfield  {author} {\bibinfo {author} {\bibfnamefont {V.}~\bibnamefont
  {Delaubert}}, \bibinfo {author} {\bibfnamefont {N.}~\bibnamefont {Treps}},
  \bibinfo {author} {\bibfnamefont {M.}~\bibnamefont {Lassen}}, \bibinfo
  {author} {\bibfnamefont {C.~C.}\ \bibnamefont {Harb}}, \bibinfo {author}
  {\bibfnamefont {C.}~\bibnamefont {Fabre}}, \bibinfo {author} {\bibfnamefont
  {P.~K.}\ \bibnamefont {Lam}},\ and\ \bibinfo {author} {\bibfnamefont {H.-A.}\
  \bibnamefont {Bachor}},\ }\bibfield  {title} {\bibinfo {title}
  {${\mathrm{tem}}_{10}$ homodyne detection as an optimal small-displacement
  and tilt-measurement scheme},\ }\href
  {https://doi.org/10.1103/PhysRevA.74.053823} {\bibfield  {journal} {\bibinfo
  {journal} {Phys. Rev. A}\ }\textbf {\bibinfo {volume} {74}},\ \bibinfo
  {pages} {053823} (\bibinfo {year} {2006})}\BibitemShut {NoStop}%
\bibitem [{\citenamefont {Sun}\ \emph {et~al.}(2014)\citenamefont {Sun},
  \citenamefont {Liu}, \citenamefont {Liu}, \citenamefont {Guo}, \citenamefont
  {Zhang},\ and\ \citenamefont {Gao}}]{doi:10.1063/1.4869819}%
  \BibitemOpen
  \bibfield  {author} {\bibinfo {author} {\bibfnamefont {H.}~\bibnamefont
  {Sun}}, \bibinfo {author} {\bibfnamefont {K.}~\bibnamefont {Liu}}, \bibinfo
  {author} {\bibfnamefont {Z.}~\bibnamefont {Liu}}, \bibinfo {author}
  {\bibfnamefont {P.}~\bibnamefont {Guo}}, \bibinfo {author} {\bibfnamefont
  {J.}~\bibnamefont {Zhang}},\ and\ \bibinfo {author} {\bibfnamefont
  {J.}~\bibnamefont {Gao}},\ }\bibfield  {title} {\bibinfo {title}
  {Small-displacement measurements using high-order hermite-gauss modes},\
  }\href {https://doi.org/10.1063/1.4869819} {\bibfield  {journal} {\bibinfo
  {journal} {Applied Physics Letters}\ }\textbf {\bibinfo {volume} {104}},\
  \bibinfo {pages} {121908} (\bibinfo {year} {2014})}\BibitemShut {NoStop}%
\bibitem [{\citenamefont {Matsumoto}(2002)}]{Matsumoto_2002}%
  \BibitemOpen
  \bibfield  {author} {\bibinfo {author} {\bibfnamefont {K.}~\bibnamefont
  {Matsumoto}},\ }\bibfield  {title} {\bibinfo {title} {A new approach to the
  cram{\'{e}}r-rao-type bound of the pure-state model},\ }\href
  {https://doi.org/10.1088/0305-4470/35/13/307} {\bibfield  {journal} {\bibinfo
   {journal} {Journal of Physics A: Mathematical and General}\ }\textbf
  {\bibinfo {volume} {35}},\ \bibinfo {pages} {3111} (\bibinfo {year}
  {2002})}\BibitemShut {NoStop}%
\bibitem [{\citenamefont {{Yang}}\ \emph {et~al.}(2018)\citenamefont {{Yang}},
  \citenamefont {{Pang}}, \citenamefont {{Zhou}},\ and\ \citenamefont
  {{Jordan}}}]{2018arXiv180607337Y}%
  \BibitemOpen
  \bibfield  {author} {\bibinfo {author} {\bibfnamefont {J.}~\bibnamefont
  {{Yang}}}, \bibinfo {author} {\bibfnamefont {S.}~\bibnamefont {{Pang}}},
  \bibinfo {author} {\bibfnamefont {Y.}~\bibnamefont {{Zhou}}},\ and\ \bibinfo
  {author} {\bibfnamefont {A.~N.}\ \bibnamefont {{Jordan}}},\ }\bibfield
  {title} {\bibinfo {title} {{Optimal measurements for quantum multi-parameter
  estimation with general states}},\ }\href@noop {} {\bibfield  {journal}
  {\bibinfo  {journal} {arXiv e-prints}\ ,\ \bibinfo {eid} {arXiv:1806.07337}}
  (\bibinfo {year} {2018})},\ \Eprint {https://arxiv.org/abs/1806.07337}
  {arXiv:1806.07337 [quant-ph]} \BibitemShut {NoStop}%
\bibitem [{\citenamefont {Padgett}(2017)}]{Padgett:17}%
  \BibitemOpen
  \bibfield  {author} {\bibinfo {author} {\bibfnamefont {M.~J.}\ \bibnamefont
  {Padgett}},\ }\bibfield  {title} {\bibinfo {title} {Orbital angular momentum
  25 years on},\ }\href {https://doi.org/10.1364/OE.25.011265} {\bibfield
  {journal} {\bibinfo  {journal} {Opt. Express}\ }\textbf {\bibinfo {volume}
  {25}},\ \bibinfo {pages} {11265} (\bibinfo {year} {2017})}\BibitemShut
  {NoStop}%
\bibitem [{\citenamefont {Maga\~na Loaiza}\ \emph {et~al.}(2014)\citenamefont
  {Maga\~na Loaiza}, \citenamefont {Mirhosseini}, \citenamefont {Rodenburg},\
  and\ \citenamefont {Boyd}}]{PhysRevLett.112.200401}%
  \BibitemOpen
  \bibfield  {author} {\bibinfo {author} {\bibfnamefont {O.~S.}\ \bibnamefont
  {Maga\~na Loaiza}}, \bibinfo {author} {\bibfnamefont {M.}~\bibnamefont
  {Mirhosseini}}, \bibinfo {author} {\bibfnamefont {B.}~\bibnamefont
  {Rodenburg}},\ and\ \bibinfo {author} {\bibfnamefont {R.~W.}\ \bibnamefont
  {Boyd}},\ }\bibfield  {title} {\bibinfo {title} {Amplification of angular
  rotations using weak measurements},\ }\href
  {https://doi.org/10.1103/PhysRevLett.112.200401} {\bibfield  {journal}
  {\bibinfo  {journal} {Phys. Rev. Lett.}\ }\textbf {\bibinfo {volume} {112}},\
  \bibinfo {pages} {200401} (\bibinfo {year} {2014})}\BibitemShut {NoStop}%
\end{thebibliography}%

\end{document}